%% file: paper.tex
\documentclass[preprint,authoryear]{elsarticle}

\input{config.tex}

\usepackage{textcomp}
\journal{Safety Science}

\newcommand{\cmmnt}[1]{}
\begin{document}

%=============================================================================
\begin{frontmatter}
\title{Modeling and mitigation of occupational safety risks in dynamic industrial environments}

%% LEAVE out for double-blind review process
\author{Ashutosh~Tewari}%\ead{ashutosh.tewari@exxonmobil.com}
\author{Antonio~R.~Paiva}%\ead{antonio.paiva@exxonmobil.com}
\address{Corporate Strategic Research, ExxonMobil Research and Engineering, Annandale, NJ, USA}

\begin{abstract}
Identifying and mitigating safety risks is paramount in a number of industries. In addition to guidelines and best practices, many industries already have \emph{safety management systems} (SMSs) designed to monitor and reinforce good safety behaviors. The analytic capabilities to analyze the data acquired through such systems, however, are still lacking in terms of their ability to robustly quantify risks posed by various occupational hazards. Moreover, best practices and modern SMSs are unable to account for dynamically evolving environments/behavioral characteristics commonly found in many industrial settings. This article proposes a method to address these issues by enabling continuous and quantitative assessment of safety risks in a data-driven manner. The backbone of our method is an intuitive hierarchical probabilistic model that explains sparse and noisy safety data collected by a typical SMS.  A fully Bayesian approach is developed to calibrate this model from safety data in an online fashion. Thereafter, the calibrated model holds necessary information that serves to characterize risk posed by different safety hazards. Additionally, the proposed model can be leveraged for automated decision making, for instance solving \emph{resource allocation problems} \textemdash targeted towards risk mitigation\textemdash \ that are often encountered in resource-constrained industrial environments. The methodology is rigorously validated on a simulated test-bed and its scalability is demonstrated on real data from large maintenance projects at a petrochemical plant.
\end{abstract}

% TODO
% %%Graphical abstract
% \begin{graphicalabstract}
% %  \includegraphics{grabs}
% \end{graphicalabstract}
%
% %%%Research highlights
% \begin{highlights}
%   \item Research highlight 1
%   \item Research highlight 2
% \end{highlights}
%

\end{frontmatter}
%\modulolinenumbers[5]
%\linenumbers

%=============================================================================

\section{Introduction}\label{sec:intro}

% safety management systems
Continuous improvement of safety performance is crucial in the construction, manufacturing, processing and other heavy industries. There have been commensurate improvements, over the past decades, with the introduction of \emph{safety management systems} (SMSs) which help to identify, assess, and mitigate occupational safety hazards. In spite of these efforts, high incident rates still plague many of these industries. This is well evidenced, for example, in the construction industry, which remains hazardous and still records a high injury rate~\citep{ringen2006construction, pinto2011}. Most crucially, these industries are seeing a plateauing trend in the improvements of safety performance~\citep{dekker2016examining}. It has been argued that these industries have reached a significant degree of saturation with respect to traditional safety strategies~\citep{esmaeili2012diffusion}. A contributing factor is that the critical risk-mitigation decisions are still predominantly made by humans, which is often ``fraught with numerous biases and misconceptions inherent to human cognition and compounds the likelihood of misdiagnosing the riskiness of a situation"~\citep{tixier2017riskModeling}. Therefore, it is imperative to consider novel approaches that objectively assess safety risks and optimally direct risk-mitigation efforts.

% new data technologies
With the digitalization of heavy industries, there have been proposals to leverage modern sensing and information management systems to breakaway from plateauing trends of safety performance. For instance, several of these proposals pertain to the tracking of persons and equipment (see \citet{awolusi2018wearable} or \citet{desvignes2014}), which can then enable monitoring of hazardous situations. For example, when someone approaches a hazardous or restricted area, an alert can be automatically generated and broadcasted ~\citep{chae2010application, arslan2020identifying}. Such new technologies have a great potential to improve workplace safety by directly monitoring the interaction between humans and their work environments. However, they are primarily used in isolation and hardly make use of the safety data collected by modern SMSs (safety observations and incidents reports), thus lacking the context about evolving safety risks~\citep{dekker2016drift, neal2004safety}. Just by themselves, these technologies provide only a highly localized and  limited perspective of the overall landscape of safety risks . Therefore, there is a need for analytic methods that provide an holistic assessment of safety risks from recorded safety data, which may then be augmented using the data from modern sensing systems.

There has been an increasing number of studies in recent years exploring the use of data analytics and machine learning techniques to identify leading indicators of safety incidents. A common solution approach in these studies is to first identify certain \emph{universal} safety attributes  (for example, using natural language processing tools to mine textual descriptions in safety reports)~\citep{tixier2016automated, zou201766, goh2017construction}, followed by the application of machine learning techniques to learn their relationship with the occurrence and the severity of actual safety incidents~\citep{sarkar2020predicting, poh2018375, tixier2016application, verma2018data, cheng2013applying}. Such studies offer the benefit of being data-driven and, when calibrated on historical data, allow one to identify attributes that are most promising leading indicators of safety incidents. In spite of these efforts, there remains a significant room for improvement in existing safety analytics approaches.
First, the aforementioned machine learning models are usually oblivious to the diverse set of challenges particularly encountered in analyzing safety data. While convenient to implement and calibrate on data, they lack the mechanisms to easily include a~priori knowledge, which may either come from safety experts or from data acquired in disparate and yet related contexts such as different projects or work-sites. Second, many of the existing analytical approaches are designed to work with \emph{reactive} data consisting of historical incident reports and severity levels. A recent trend in SMSs is to gather more \emph{proactive} data in the form of safety observations/inspections~\citep{verma2018data}, to supplement the often sparse (fortunately) reactive data. However, there are no well-established approaches that combine the learnings from both reactive and proactive data in a principled and meaningful manner. Lastly, while the existing methods are helpful, their impact is limited to a snapshot in time since they are not designed to be attuned to dynamic work environments with evolving work-types and safety risks. 

The proposed framework is similar to the existing ones in certain aspects, but there are several points of departure that help to overcome the aforementioned drawbacks. To be specific, the contribution of this study is threefold. \textbf{First}, a hierarchical probabilistic model of the safety data is proposed that ties in the generation of reactive and proactive data, thus providing a common platform to assess risk of safety hazards from different data types.  Safety data records are temporally sparse and obtained at irregular intervals. Within the theory of random processes, such data is better characterized as point processes~\citep{daley1988introduction}. Unlike the existing approaches of safety risk assessment, ours takes advantage of this viewpoint when analyzing such data. \textbf{Second}, an efficient, recursive Bayesian inference algorithm is developed to calibrate the aforementioned probabilistic model from the safety data in an online fashion. As a result, the model evolves with the changing work environment, thereby providing  the ability of dynamic risk assessment. \textbf{Third}, while the assessed risks are useful in their own right (by helping safety professional identify areas to intervene upon), we demonstrate the applicability of our approach for optimal decision making for a particular resource allocation problem.

\emph{Remark}: Due to the focus of this study on continuous risk assessment from safety observations and reports, our methodology is arguably focused on \emph{behavioral} aspects of safety. These aspects tend to be more dynamic and evolve more closely in interdependence with safety culture, attitudes, and other exogenous conditions~\citep{dekker2016drift, curcuruto2015role, fugas2012468, wu2011716}. In contrast, systemic risks stemming from processes and equipments are usually effectively mitigated by designing appropriate safeguards and best practices, and hence are out of scope of this study. \cmmnt{Yet, human factors remain and must be continuously monitored and conditioned~\citep{dekker2016drift}.}

The remainder of this paper is organized as follows. First, in \secref{sec:background} we describe the general characteristics of a SMS assumed in this work. This includes the interplay between proactive and reactive safety data that needs to be available, and the detail therein, such that it enables the assessment of safety risk. A formal problem statement is provided in \secref{sec:problem}. The proposed analytical methodology is then presented in \secref{sec:method}, comprising the probabilistic model for dynamic risk assessment in \secref{subsec:Model}, the recursive Bayesian estimation of this model in \secref{subsec:Estimation}, and the optimization strategy for a specific resource allocation problem in \secref{subsec:Optimization}. The methodology is validated in a simulated environment and demonstrated for its scalability in a real-world setting in \secref{sec:results}. Conclusions and considerations for future work are presented in \secref{sec:discussion}.

\section{Background on Safety Management Systems (SMSs) }\label{sec:background}

Over the last decade or two, the introduction of SMSs has proven to be crucial in the improvement of workplace safety. First and foremost, these systems provide a systematic process to record safety data, which then help identify high-risk elements, which in turn leads to best practices and safeguards to mitigate them. Additionally, they have also brought to the forefront the interdependence between safety climate and safety behavior, and their ultimate effect on safety performance~\citep{fugas2012468, wu2008correlation, wu2011716}. For example, a workforce general attitude toward safety is undoubtedly essential in the collection of the necessary data and the translation of recommendations into improved safety practices and behavior. This means that the success of a safety program is underpinned by the appropriate safety climate and the necessary leadership support~\citep{sheehan2016leading, fugas2012468}. 

The risk modeling and mitigation methodology described in this study works within the framework of a general workplace SMS. There are several characteristics that must be present that we describe in this section, and also provide the rationale behind them. First of all, there must exist a substantial focus on collecting proactive safety data, typically in the form of safety observations. These not only help to dynamically assess safety risk but also serve to reinforce safety awareness and safe practices. Moreover, both \emph{positive} and \emph{negative} observations are needed~\citep{grant2018back} (refer to \secref{sec:observations}). Secondly, near-misses and incident report should attempt to characterize the \emph{potential} injury severity and not only the actual severity incurred (see \secref{sec:hurt.level} for details). As will be shown, this allows for an improved assessment of safety risk. It is worth mentioning that although one could use the proposed methodology in a SMS without these characteristics, its usefulness would be limited due to the sparsity of information available through actual incidents.

\subsection{Safety data: observations and incidents}\label{sec:observations}

As previously mentioned, the data collected by modern SMSs primarily consists of safety observations and near-miss \& incident reports. Safety observations can be considered as a proactive form of data that includes assessment reports on workers' behaviors and workplace conditions. These observations are not contingent upon safety incidents and are made regardless of them. Safety observations should capture both safe (i.e., \emph{positive}) circumstances and at-risk or unsafe (i.e., \emph{negative}) circumstances so as to appropriately characterize the relative frequency of these situations~\citep{grant2018back, hollnagel2012fram}.
It is a common practice to group these observations into categories pertaining to work types, areas of focus, or different hazards~\citep{verma2018data, tixier2017riskModeling}. In the remainder of the paper we refer to such categories as \emph{vulnerabilities} that, in our opinion, is a more appropriate term in the context of workplace safety.

% observations
In this study, we consider three types of safety observations: work-safety observations~(WSOs), safety assessment observations~(SAOs), and best-practices observations~(BPOs). WSOs are general observations collected by workers and leadership capturing both safe and unsafe behaviors, practices, or situations. For example, a WSO might be submitted by a worker reflecting on how a colleague proceeded when realizing that it was lacking the proper personal protective equipment~(PPE) for a task, or when a worker noticed a workplace hazard because of cabling across a path or a misplaced barrier. In both cases the observations are associated with appropriate vulnerabilities, for instance, ``PPE" for the former and ``Barricades" for the latter.   SAOs are more formal observations in the sense that they are collected by designated safety inspectors by actively observing workers as they perform their tasks. BPOs, like SAOs, are also deliberate observations albeit focused on the execution or critical evaluation of best-practices for well-defined tasks. Like WSOs, SAOs and BPOs are associated with the vulnerabilities that become apparent during scrutiny. Because of the more deliberate nature of SAOs and BPO, they may be given higher weight when assessing the risk of vulnerabilities. On the other hand, one may argue that SAOs and BPOs can be primed towards observing more negative behaviors/situations. Regardless of the observation types, it is logical to assume that such observations are subject to biases (because of humans in the loop) and knowing the appropriate weights of each observation type is nontrivial. Our approach addresses this gap by allowing to infer weights of each observation types in a data-driven manner.

% near-misses and incident reports
While observations reflect the general safety climate, assessing safety risk requires inferring how those observations relate to the safety incidents and near-misses. The latter is especially important because it could be a crucial leading indicator of a future incident. In the literature the importance of near-miss reports in identifying, assessing, and controlling idiosyncratic risks has also been repeatedly emphasized~\citep{haas2020learning, yorio2014interpreting, lander2011near}. In this study we consider near-misses as yet another form of incidents. In fact, it can be argued that near-misses differ from incidents only in the severity of the actual outcome. By taking into consideration an assessment of the severity potential (\secref{sec:hurt.level}) near-misses are indistinguishable and play a central role in assessing the overall safety risk. Therefore, in the remainder of this article ``incidents'' should be considered to include both near-misses and actual incidents. Lastly, like safety observations, incidents are associated with appropriately chosen vulnerabilities.  

\subsection{Incident severity (actual and potential)}\label{sec:hurt.level}
% HURT system; AHL and PHL

The severity of an injury associated with an incident provides only partial information regarding the state (or riskiness) of a vulnerability. In reality there is substantial variability in the injury severity of incidents, including no injury in the case of near-misses. Accordingly, each incident provides an opportunity to assess the distribution of injury severities for the vulnerability in consideration.  \cmmnt{Traditionally, severities of personal injury incidents are quantified according to a treatment-based approach, which reflects the type of medical treatment used in response. While this approach is common among many industries and often required in regulatory compliance, it emphasizes administrative reporting and incident escalation management procedures. As a result, low-severity injuries tend to be overlooked and this fails to convey a safety culture of caring and improvement. Moreover, treatment-based metrics focus on the severity of past incidents, and therefore are lagging indicators of future incidents. Perhaps most importantly, this approach does not include an assessment of potential (worst-case) consequences, which are critical to understand and prevent future incidents.} To this end, we adopt a \emph{Hurt-based} approach that has been recently proposed~\citep{smith2014hurt, etaje2013efficiency}. This approach provides a framework to consistently describe an actual injury severity and evaluate the potential (worst-case) injury severity. Moreover, this approach can be extended to qualify the outcome severity of process, property, and environmental incidents, thereby providing the means to effectively integrate learnings from common potential effects across incident types~\citep{eggleston2014assessing}. For these reasons, the Hurt-based approach is assumed in the remainder of this paper. To be specific, the modeling method proposed in \secref{sec:method} shall assume that incidents contain an evaluation of both actual and potential injury severities. In the following, AHL (actual Hurt level) and PHL (potential Hurt level) will be used to denote the actual and potential severity of an injury following an incident, respectively. They take values along 6 levels, zero through five. A Hurt-level of zero means that no injury took place, meaning that it was a near-miss (or property or environmental incident, if considered). Intuitively, the severity rapidly escalates with increasing levels, with levels 4 and 5 corresponding to one and multiple fatalities, respectively. By definition, the PHL is always equal or greater than the AHL.

Although the modeling herein assumes that the Hurt-based approach was employed in an SMS, the model could be easily modified to work with traditional severity metrics. This is equivalent to limiting oneself to using only the AHL. This is not ideal, as discussed later, since the evaluation of PHL can be crucial for inferring the distribution of severity outcomes, and thus for accurate assessment of future safety risk.

% TODO: grouping of concern areas as *vulnerabilities*

\input{problem_setup.tex}

\input{method.tex}

\input{results.tex}

%\begin{figure}
%  \centering
%  \includegraphics[width=0.48\textwidth]{}
%  \caption{.}
%  \label{fig:}
%\end{figure}

\section{Discussion and Conclusion}\label{sec:discussion}
This paper presents a method to quantify occupational safety risks, a task central for workplace safety in heavy industries such as construction, manufacturing and processing. The method is grounded in probability theory and Bayesian statistics, and is designed to account for real-world issues that include sparse information, biased observations, dynamically evolving workplace conditions among others. By virtue of the Bayesian formalism, our modeling framework provides a systematic way of incorporating prior knowledge from safety experts and reconciling it with the safety data. Additionally, it enables us to obtain various risk measures (e.g. expected-loss or tail-probability) for all safety hazards (or vulnerabilities), which can then be monitored in real-time. Such monitoring can be immensely helpful \textemdash for instance generating a seriatim of hazards by risk\textemdash for a safety manager to make informed risk-mitigation decisions. The method is primarily intended to enhance human judgment by identifying potential blind-spots which can arise in a dynamic industrial work environment involving various occupational hazards. Nevertheless, we also outline an approach to address a commonly encountered problem of optimal resource allocation that organically leverages the proposed  risk assessment model in a stochastic optimization framework. We rigorously demonstrate  the effectiveness of risk-based resource allocation method over a heuristic method, representative of approaches often used in practice for their simplicity. For completeness, we also apply the proposed risk assessment model to an actual industrial problem to show its ability to process noisy, real-world data while yielding consistent and meaningful risk assessments.

Our attempt to bring the same mathematical rigor \textemdash in the assessment of safety risks \textemdash that is found in other areas such as seismology \citep{cornell1968}, actuarial science \citep{vose2008} and epidemiology \citep{lawson2018} has primarily two motivations. The first, of course, is to equip managers with tools sophisticated enough to meaningfully analyze sparse safety data and provide actionable insights. The second, perhaps equally important, is to understand the value of information in different types of safety data being collected and identify any gaps therein. In that regard, one has little control over reactive data (actual incidents and injury severity levels), leaving us with the question that \emph{how the quality of proactive data (safety field observations) can be improved so that it enhances our ability to assess and mitigate risks of safety hazards}? We believe that the modeling framework presented in this work can help answer such questions. Having said that, the framework itself can be extended on several fronts and can be tailor-made for a particular work-site. For instance, a simplifying assumption made throughout is that all safety hazards are independent of each other. One may argue against this assumption by citing examples where increased riskiness in one hazard was indicative of increased risks in others. To that end, the proposed statistical model can be modified to incorporate such dependencies via additional parameters. However, how faithfully such parameters can be inferred would greatly depend on the quality of safety data being collected. There may be other aspects of the proposed framework that could invoke questions as we transition from theory to practice. Our hope is that this work (or any derived work) provides a systematic mechanism to answer such questions with the overarching goal of improving workplace safety across all industries.

\cmmnt{ \section{Acknowledgments}
The authors would like to thank the following people for their involvement in different phases of this work. Loan Tran and Joseph Covington for providing initial guidance on the challenges and opportunities for safety risk management in a petrochemical plant, Matthew Cobb for consultation on how the identified opportunities can be realized and made practical, Jose Hernandez and Chuck Knight for providing IT support to acquire real-world data used in this study, and Nathaniel Rogalskyj for helpful discussions on the development of the environment simulator.  We would also like to thank Paul Austin for his careful perusal of the manuscript as an expert in the area of industrial safety. Finally, we would like to thank ExxonMobil's management for their continued support to advance the critical area of industrial safety risk management. }

%-------------------------------------------------------------
%\clearpage
\bibliographystyle{elsarticle-harv}
\bibliography{refs}

\end{document}

%% file: config.tex
\usepackage{textcomp}
\usepackage{soul}
\usepackage{comment}
\usepackage{amssymb,amsmath,amsfonts,bm,mathrsfs}
\usepackage{graphicx,stfloats}
\graphicspath{{./}{figs/}}
\usepackage{float}
\usepackage{subfig}

\usepackage{cite,url,enumerate,xspace}
\usepackage{array,multirow,dcolumn,booktabs,longtable,xcolor}
\usepackage{standalone}
\usepackage{natbib}
\usepackage[hmargin=1.25in, vmargin=1in]{geometry}
\usepackage{setspace}
\newcommand{\PreserveBackslash}[1]{\let\temp=\\#1\let\\=\temp}
\newcolumntype{C}[1]{>{\PreserveBackslash\centering}m{#1}}
\newcolumntype{d}[1]{D{.}{.}{#1}}

\newcommand{\x}{\+x}

\newcommand{\E}[2][]{\ensuremath{%
  \if\relax#1\relax\mathbb{E}\!\left[#2\right]
  \else\mathbb{E}_{#1}\left[#2\right]\fi}}

\newcommand{\equationref}[1]{eq.~\eqref{#1}}
\newcommand{\figref}[1]{figure~\ref{#1}}

\newcommand{\figsref}[1]{figures~\ref{#1}}

\newcommand{\secref}[1]{Section~\ref{#1}}

\newcommand{\incidents}{\textbf{e }}

\newcommand{\safetydata}{\ensuremath{\mathcal{D}[t]\,}}
\newcommand{\uobs}[1] {a^-_{{#1}}}

\newcommand{\prior}{\pi_{\text{pr}}}
\newcommand{\like}{\mathcal{L}}

\DeclareMathOperator*{\argmin}{argmin}

\DeclareMathAlphabet{\mymathbb}{U}{BOONDOX-ds}{m}{n}

\makeatletter
\def\fps@figure{tbp}
\def\fps@table{tbp}
\makeatother

%% file: problem_setup.tex
\section{Problem description}\label{sec:problem}
Consider an industrial work environment equipped with a SMS that records safety data, of the form discussed in Sections~\ref{sec:observations} and \ref{sec:hurt.level}, on a daily basis. We assume that any task in this work environment is fully described by a set of $n$ unique vulnerabilities.  Let \safetydata denote the safety data recorded at time $t$. This includes the number of incidents, the associated Hurt levels ($\textbf{ahl}$, $\textbf{phl}$) and the number of safety observations made of various types (WSOs, BPOs, etc.).  Thus, \safetydata is a collection of data recorded for all the $n$ vulnerabilities at time $t$; that is,
\begin{align} \label{eq:FieldData}
  \safetydata = \left\{ n^i_\incidents[t],\mathbf{ahl}^i[t],\mathbf{phl}^i[t], n^i_{a_{X_1}}[t], n^i_{a_{X_2}}[t], n^i_{\uobs{X_1}}[t], n^i_{\uobs{X_2}}[t] \right \}_{i=1}^n
\end{align}
where, for the $i^{th}$ vulnerability, $n^i_\incidents[t], \mathbf{ahl}^i[t], \mathbf{phl}^i[t]$ are the number of incidents and the vectors of corresponding AHLs and PHL values, respectively. Note that the length of vectors  $\mathbf{ahl}^i[t]$ and $\mathbf{phl}^i[t]$ is $n^i_\incidents[t]$.  Similarly $n^i_{a_{X_*}}[t] \text{ and } n^i_{\uobs{X_*}}[t]$ denote the number of total and the negative observations recorded for any observation type ($X_*$), respectively. Here we restrict to two observation types denoted as $X_1$ and $X_2$ for the ease of exposition, but the proposed framework works for an arbitrary number of observation types. Additionally, let's assume that the total number of observations of all types, across all vulnerabilities and at all time-steps is finite and is denoted by $m$. In other words, fixed number of observations are made every day due to finite resources. Given this data, we have two main goals as follows.
\begin{itemize}
 \item Continuously monitor the \emph{state} of all vulnerabilities based on safety data. The state of the $i^{th}$ vulnerability is specified by two distribution viz. $p(n_\incidents^i)$ and $p(\textbf{ahl}^i|\ n_\incidents^i \neq 0)$. The former characterizes the incidence occurrence rate and the latter, the injury severity if an incident occurs. These two distributions are sufficient to yield a meaningful measure of risk for each vulnerability as discussed shortly in \secref{sec:risk}.
\item Given the current state of all vulnerabilities, allocate $m$ observations across $n$ vulnerabilities. Safety observations are onerous (requires human in the loop) and scarce compared to the size of the workforce and the number of tasks being carried out in any given day. Therefore, a SMS may greatly benefit by making these observations judiciously i.e. more resources are allocated towards vulnerabilities with higher risks.  
\end{itemize}
Note that it might just be sufficient to quantify the risks associated with each vulnerability (first of the two goals), and present that information to the leadership to act upon it, for instance use this information to allocate observation resources. Nevertheless, our second goal attempts to mathematically formalize that process as well for scenarios where the work environment is just too large with many moving parts, and any manual decision making is not just cumbersome but also susceptible to human error in judgment. 

\subsection{Safety risk}\label{sec:risk}
% discussion of "safety risk": incident likelihood vs severity
Safety risk is usually evaluated along two main dimensions, i.e. the incidents frequency and their severity ~\citep{baradan2006}. On one hand, the number of incidents, of course, tells us how often injuries are occurring, and is a reflection of the state of overall safety landscape. On the other hand, it provides an incomplete picture without the information about the incidents' severity. In the majority of situations it is critical to avoid the occurrence of high-severity incidents, meaning that, while one may want to eliminate minor injuries, such as small cuts or scrapes, those are of drastically lower importance relative to the prevention of life-altering injuries or fatalities. Therefore, considering both dimensions is undoubtedly important to characterize safety risk.

A logical way to define safety risk is via the product of frequency and severity of incidents for every vulnerability~\citep{tixier2017riskModeling}. We adhere to the same logic but define risk in probabilistic terms.  Let $p(\mathbb{C}^i_{ahl=j})$ denote the distribution of daily counts of incidents of the Hurt level $j$ for the $i^{th}$ vulnerability. \secref{subsec:Optimization} details how this distribution can be obtained after calibration with the historical safety data.  Associating a loss (or severity score) with each Hurt-level, a loss vector can be defined as $\mathbf{c} = [0,1,10,100,1000,10000]^T$. Here the associated loss values (and units) are rather arbitrary but capture an essential feature that with each increment in the Hurt-level the corresponding increase in the loss is exponential. An alternate set of severity scores can be found elsewhere~\citep{hallowell2010}, exhibiting similar exponential behavior.  With these two pieces of information the risk can be defined, for each vulnerability, as the \emph{expected-loss} i.e. $s^i = \sum_{j=0}^5\mathbf{c}[j] \cdot\mathbb{E}(\mathbb{C}^i_{ahl=j})$, where $\mathbb{E}$ is the expectation operator with respect  to the distribution $p(\mathbb{C}^i_{ahl=j})$.  \secref{sec:method} presents an in-depth exposition of how the risk can be computed and updated in an online fashion. 

One may argue that the definition of risk by~\citet[see Equation 2]{tixier2017riskModeling} is similar to the one defined here as the expected-loss. This argument is only partly true since there are some notable differences and advantages of the latter. First, the former obtains risk values from historical incident data. This approach, while suitable in  data rich scenarios, is susceptible to incorrect assessment of risk when the data is sparse, as is often the case with safety incidents. The proposed approach, on the contrary, relies on the distribution of Hurt levels which is learned not only from the  (scarce) incident data but also from the observation data (proactive data) and any available prior knowledge (see \secref{sec:method} for details). Thus, the characterized risk is more robust to data sparsity and noise. Secondly, having access to the Hurt level distribution $p(\textbf{ahl}^i)$, other risk measures, popular in financial risk management, such as \emph{value at risk} (VaR) and \emph{conditional value at risk} (CVaR) ~\citep{pflug2000} can be easily derived. These risk measures emphasizes the tail portion of the Hurt level distribution (rare, high severity incidents), and thus could be preferred by leadership (over the expected-loss) to make risk-mitigation decisions.

%% file: method.tex
\section{Safety Risk Management}\label{sec:method}

Safety risk management involves both the assessment of various safety risks as well as their mitigation by taking proper, targeted actions. An ideal framework for the same should be able to work with sparse safety data, while being able to account for various real world uncertainties. To this end we present an approach that is grounded on probabilistic modeling and inference, and has the following salient features. The proposed approach:
\begin{itemize}
\item characterizes risk posed by every vulnerability based on both reactive and proactive safety data via a common hierarchical probabilistic model (refer to \secref{sec:risk}),
\item works in an online fashion by maintaining the state of a vulnerability at every time point and updating it as soon as new safety data arrives,    
\item provides a systematic approach to incorporate potential Hurt level information to compensate for scarce actual Hurt level information from incidents, 
\item automatically quantifies the information content in each observation type, and
\item lends itself to stochastic optimization methods for making optimal risk-mitigation decisions.  
\end{itemize}

We first hypothesize a hierarchical probabilistic model that describes the generative process of the safety data \safetydata as described in \secref{sec:problem}. Once calibrated on the safety data, this generative model forms the foundation of our risk management framework. We first describe the proposed generative model in \secref{subsec:Model} and provide the intuition behind it. This model involves a few free parameters, for which a fully Bayesian parameter estimation procedure presented in \secref{subsec:Estimation} can be implemented in an \textit{online} fashion. Lastly in \secref{subsec:Optimization}, we define the risk scores for every vulnerability  that can be easily computed from the calibrated generative model. In addition, we provide details of a stochastic optimization framework to solve a resource allocation problem aiming to minimizes cumulative future risk. 

\subsection{Model}\label{subsec:Model}
In the discussion henceforth we omit the time index $[t]$ for the readability of equations. The time index will be reintroduced in the equations when appropriate.   The governing equations (\equationref{eq:genproc_X1} -- \equationref{eq:genproc_PHL}) completely specify the generative model for the data \safetydata as given in \equationref{eq:FieldData}.
\begin{subequations}
\begin{align}
n^i_{\uobs{X_1}} &\sim \like(n^i_{\uobs{X_1}})=\text{Binomial}(n^i_{\uobs{X_1}}|n^i_{a_{X_1}},\kappa^i_{X_1})\label{eq:genproc_X1} \\
n^i_{\uobs{X_2}} &\sim \like(n^i_{\uobs{X_2}}) =\text{Binomial}(n^i_{\uobs{X_2}}|n^i_{a_{X_2}},\kappa^i_{X_2}) \label{eq:genproc_X2}\\
n^i_\incidents &\sim \like(n^i_\incidents) =\text{Poisson}(n^i_\incidents | \kappa^i \beta^i); \text{ s.t. } \kappa^i = w_{X_1}\kappa^i_{X_1}+w_{X_2}\kappa^i_{X_2}  \label{eq:genproc_incidents}  \\
\mathbf{ahl}^i &\sim \like(\mathbf{ahl}^i) = \text{Multinomial}(\mathbf{ahl}^i |n^i_\incidents,\mathbf{p}^i) \label{eq:genproc_AHL} \\
\mathbf{phl}^i &\sim \like(\mathbf{phl}^i) =\text{Multinomial}(\mathbf{phl}^i |n^i_\incidents,\mathbf{p}^i,\mathbf{ahl}^i) \label{eq:genproc_PHL} 
\end{align}
\end{subequations}
The parameter set $\Theta = \left[\left\{\beta^i,\kappa^i_{X_1},\kappa^i_{X_2},\mathbf{p}^i\right\}_{i=1}^n,w_{X_1},w_{X_2} \right]$ characterizes this generative model completely.  Equation~\eqref{eq:genproc_X1} (or \equationref{eq:genproc_X2}) models the number of unsafe acts recorded from observations of type $X_1$ (or $X_2$) as a \textit{Binomial process} with unknown failure probability $\kappa^i_{X_1}$  (or $\kappa^i_{X_2}$). Thus, the parameters $\kappa^i_{X_1}$  (or $\kappa^i_{X_2}$) can be interpreted as the proportion of tasks, corresponding to the $i^{th}$ vulnerability, that are likely to be flagged as unsafe by the observation type $X_1$ (or $X_2$). Since, the observation processes may be biased, these parameters are, at best, noisy estimates of the actual proportion of the tasks that are being carried out unsafely. Hence, an improved estimate ($\kappa^i$) of the same is obtained as a weighted combination of $\kappa^i_{X_1}$ and $\kappa^i_{X_1}$ (refer to \equationref{eq:genproc_incidents}). The weights over observation types ($w_{X_1}$ and $w_{X_2}$) are assumed to be positive and sum to 1. Equation~\eqref{eq:genproc_incidents} posits that the number of a incidents is a \textit{Poisson process} with rate parameter $\kappa^i \times \beta^i$. The $\kappa^i$ term ties the proportion of unsafe tasks to the occurrence of incidents, in the sense that the higher the proportion of unsafe tasks the higher the incident rate. This is a key feature of our method because it provides a mechanism to \emph{explicitly} tie proactive (safety observations) and reactive (safety incidents) data, which provides an intuitive and direct explanation of safety data unlike other approaches proposed so far in the literature. The $\beta^i$ term accounts for all the other unknown factors that would determine the incident occurrence rates. Finally, given the number of incidents ($n^i_\incidents$) for the $i^{th}$ vulnerability, \equationref{eq:genproc_AHL} postulates that the corresponding AHLs are distributed according to a multinomial distribution defined by the positive parameter vector $\textbf{p}^i$, which sums to 1. Certain vulnerabilities are inherently riskier (or less risky) than others, and result in incidents with higher (or lower) AHL. That information is encoded  in the parameter vector $\textbf{p}^i$. Additionally, PHLs are also distributed with the same multinomial distribution, but are conditional on the AHL values, since a PHL is always greater than or equal to the corresponding AHL.

\begin{figure}
  \includegraphics[width=1.0\textwidth]{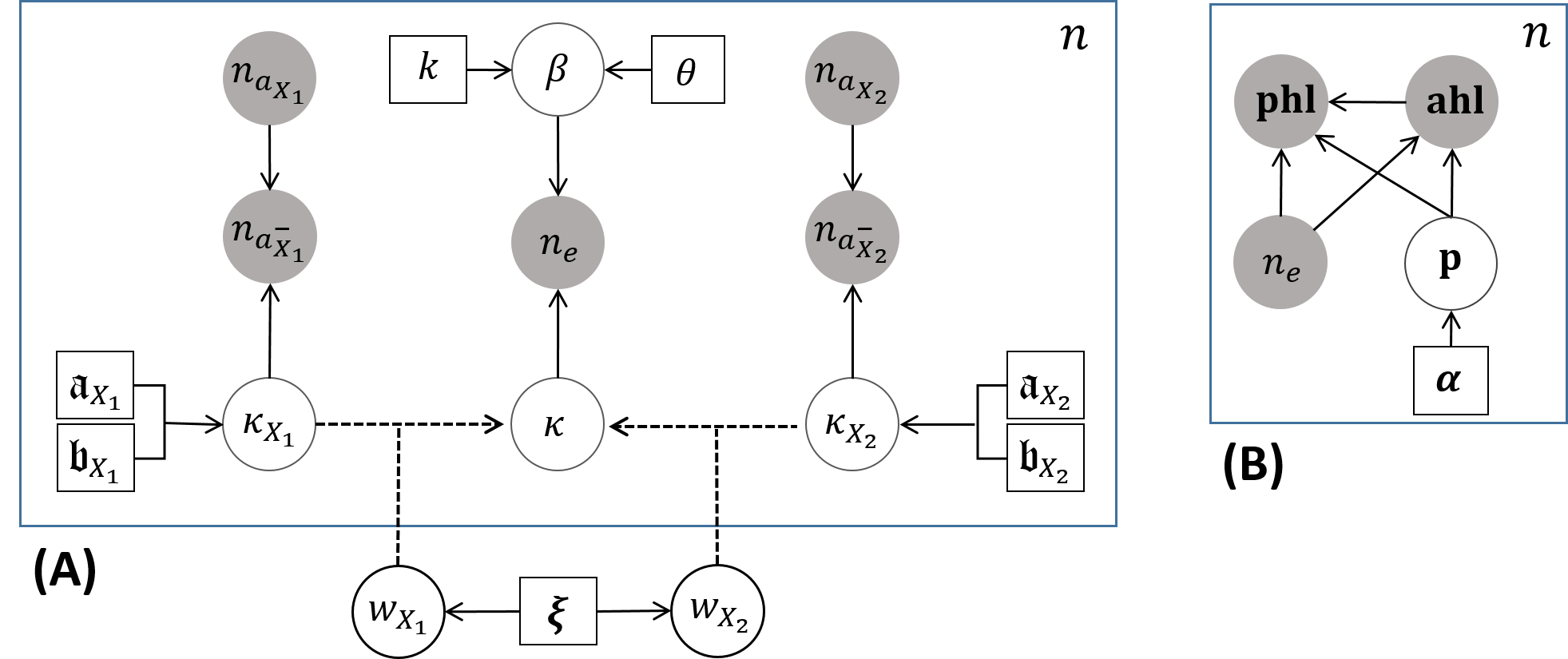}
  \caption{Illustration of the generative model described in this section using the standard \emph{plate notation} for expressing graphical models. The shaded and unshaded circular nodes represent observed  and unobserved random variables, respectively. The square nodes denote the hyper-parameters to be inferred from the safety data. The solid arrows indicate a sampling operation, while the dotted ones a deterministic relationship. The plate representation on left~(A) describes to the generative model for the event counts (incidents and observations) and corresponds to  eqs.~\eqref{eq:prior_w}--\eqref{eq:prior_gamma2} and eqs.~\eqref{eq:genproc_X1}--\eqref{eq:genproc_PHL}. The plate representation on the right~(B) corresponds to the generative model of the Hurt levels (actual and potential) associated with the incidents (eqs.~\eqref{eq:genproc_AHL}--\eqref{eq:genproc_PHL} and \equationref{eq:prior_p}). Both of these representations correspond to one time step and to one vulnerability, and applies to all of the $n$ vulnerabilities as indicated by the symbol $n$ in the top right corners. At every time step, the hyper-parameters (square nodes) are inferred from the data and the prior knowledge from the previous time step. These inferred values serve as a prior for the next-time step. This recursive Bayesian estimation is described \secref{subsec:Estimation}. }
  \label{fig:generative_model_plate_rep}
\end{figure}

To complete the hierarchical Bayesian model of the safety data, we specify  prior distributions on the unknown parameters $\Theta= \left[\left\{\beta^i,\kappa^i_{X_1},\kappa^i_{X_2},\mathbf{\alpha}^i\right\}_{i=1}^n,w_{X_1},w_{X_2} \right]$. The prior distributions have free parameters themselves (often referred to as \emph{hyper}-parameters). In such hierarchical settings, a Bayesian update via the application of Bayes rule (\equationref{eq:BayesRule}) amounts to updating these hyper-parameters as we would see in the next section.  For our problem we choose the following prior distributions for the unknown parameter set $\Theta$. 
\begin{subequations}
\begin{align}
\mathbf{p}^i &\sim \prior(\mathbf{p}^i)= \text{Dirichlet}(\mathbf{p}^i|\alpha^i) \label{eq:prior_p}\\
[w_{X_1},w_{X_2}]=\mathbf{w} &\sim \prior(\mathbf{w})= \text{Dirichlet}(\mathbf{w}|\xi) \label{eq:prior_w}\\
\beta^i &\sim \prior(\beta^i) = \text{Gamma}(\beta^i|k^i,\theta^i) \label{eq:prior_beta}\\
\kappa^i_{X_1}  &\sim \prior(\kappa^i_{X_1}) = \text{Beta}(\kappa^i_{X_1}|\mathfrak{a}^i_{X_1},\mathfrak{b}^i_{X_1}) \label{eq:prior_gamma1} \\
\kappa^i_{X_2}  &\sim \prior(\kappa^i_{X_2}) = \text{Beta}(\kappa^i_{X_2}|\mathfrak{a}^i_{X_2},\mathfrak{b}^i_{X_2}) \label{eq:prior_gamma2}
\end{align}
\end{subequations}
These are some natural choices of the prior distributions for the proposed generative model of the safety data. For instance, a commonly used prior distribution of positive vectors that sum to 1 is the Dirichlet distribution (\equationref{eq:prior_p} and  \equationref{eq:prior_w}). Similarly, proportions are typically specified using Beta distribution (\equationref{eq:prior_gamma1} and  \equationref{eq:prior_gamma2}), and positive scalars denoting intensity rate of a Poisson process as Gamma distribution (Equation \equationref{eq:prior_beta}). For the benefit of readers we include a graphical representation, in  figure \ref{fig:generative_model_plate_rep}, of the aforementioned generative model of the safety data. 

\subsection{Estimation}\label{subsec:Estimation}
We present a fully Bayesian approach to estimate the parameters of the aforementioned model in an online fashion. The \textit{Bayes} rule in \equationref{eq:BayesRule} underpins the parameter update from time $t$ to  $t+1$. Given the prior distribution $\pi_{\text{pr}}(\Theta)$ on parameters (eqs.~\eqref{eq:prior_w}--\eqref{eq:prior_gamma2}) and the data likelihood function $\mathcal{L}(\safetydata|\Theta))$ (eqs.~\eqref{eq:genproc_X1}--\eqref{eq:genproc_PHL}), the Bayes rule provides a mapping from the prior to the posterior distribution $\pi_{\text{post}}(\Theta)$.
\begin{align}\label{eq:BayesRule}
\pi_{\text{post}}(\Theta) \propto \pi_{\text{pr}}(\Theta)\mathcal{L}(\safetydata|\Theta)
\end{align}
The prior distribution at initial time step ($t=0$) may encode domain knowledge, which then gets progressively refined in subsequent time steps as new data is observed. This is achieved by the recursive application of \equationref{eq:BayesRule}, such that the posterior at the time $t$ serves as the prior for the next time step $t+1$.  Given the prior distributions (eqs.~\eqref{eq:prior_w}--\eqref{eq:prior_gamma2}) and the likelihood functions  (eqs.~\eqref{eq:genproc_X1}--\eqref{eq:genproc_PHL}), the right hand side of  \equationref{eq:BayesRule} can be written as,
\begin{align}\label{eq:unnorm_posterior}
& \pi_{\text{pr}}(\Theta)\mathcal{L}(\safetydata|\Theta) = \nonumber \\
&\prior(\mathbf{w}) \prod_{i=1}^n \prior(\mathbf{p}^i) \prior(\beta^i) \prior(\kappa^i_{X_1}) \prior(\kappa^i_{X_2})  \like(n^i_{\uobs{X_1}}) \like(n^i_{\uobs{X_2}}) \like(n^i_\mathbf{e}) \like(\mathbf{ahl}^i) \like(\mathbf{phl}^i).
\end{align}
Rearranging \equationref{eq:unnorm_posterior} by exploiting parameter independence, yields an unnormalized posterior that can be written as product of the two unnormalized posteriors as shown below.
\begin{align}
& \pi_{\text{pr}}(\Theta)\mathcal{L}(\safetydata|\Theta) = \nonumber \\
&\left[ \prior(\mathbf{w})  \prod_{i=1}^n \prior(\beta^i) \prior(\kappa^i_{X_1}) \prior(\kappa^i_{X_2})  \like(n^i_{\uobs{X_1}}) \like(n^i_{\uobs{X_2}}) \like(n^i_\mathbf{e}) \right] \left[ \prod_{i=1}^n \prior(\mathbf{p}^i)  \like(\mathbf{ahl}^i) \like(\mathbf{phl}^i) \right]
\end{align}
The latter of the two posterior admits a closed-form representation (as detailed in \secref{subsubsec:Posterior_part2}) because of the conjugacy of the prior with its likelihood function. However, for the former, one needs to approximate the posterior using sampling methods. Therefore, we propose a hybrid scheme, where at each Bayesian update the hyper-parameters $\mathbf{\alpha}^i$ that characterizes the second posterior are obtained analytically (refer to \secref{subsubsec:Posterior_part2}), while the hyper-parameters ($\xi, k^i,\theta^i, \mathfrak{a}^i_{X_1},\mathfrak{b}^i_{X_1},  \mathfrak{a}^i_{X_2}, \mathfrak{b}^i_{X_2}$) of the first posterior are relearned from samples obtained via HMC algorithm (refer to \secref{subsubsec:Posterior_part1}). The hyper-parameters so learned are then used to specify priors for the next time step. This process is repeated at every time step thereby ensuring continuous update of parameters as new safety data is acquired.

\subsubsection{Simulation of $\pi_{post}\big(\mathbf{w},\beta^i,\kappa^i_{X_1}, \kappa^i_{X_2} \ | \ n^i_{\uobs{X_1}}, n^i_{\uobs{X_2}},n^i_\mathbf{e} \big)$} \label{subsubsec:Posterior_part1}
\begin{figure}
  \centering
  \fbox{\includegraphics[width=5in, trim=0 6mm 0 0, clip]{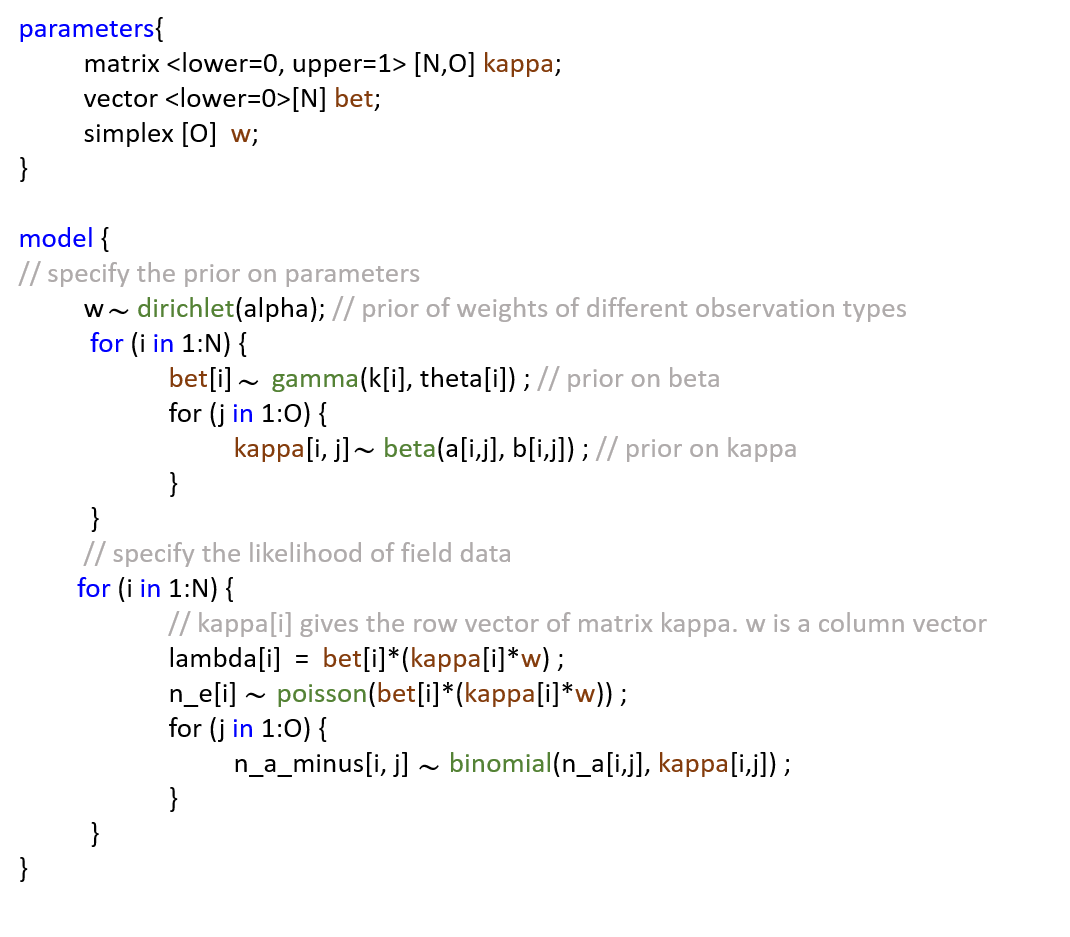}}
  \caption{Stan code implementing the proposed generative model to carry out the inference. The succinct syntax of Stan probabilistic programming is evident as one can notice direct correspondence with the generative model outlined in section \ref{subsec:Model}. The inference (MCMC or variational Bayes) is automatic and require no further effort from the modeler.}
  \label{fig:stan_code}
\end{figure}

Because of the analytical intractability of this posterior, we simulate it using the Hamiltonian Monte Carlo (HMC) algorithm using Stan~\citep{carpenter2017}, a powerful probabilistic programming language.  Recent years have seen a surge in expressive probabilistic programming languages such as Stan. The primary motivation behind such languages is to make probabilistic inference easier by offering two primary benefits. First, a succinct syntax to describe complex probabilistic models that otherwise would take many lines of code in traditional programming languages, and second, the inference is automatic and does not require any additional effort.  A snippet of Stan code that implements the proposed generative model for the safety data is shown in \figref{fig:stan_code}. The one-to-one correspondence between the mathematical representation of the generative model (Equations ~\eqref{eq:prior_w}--\eqref{eq:prior_gamma2} and Equations ~\eqref{eq:genproc_X1}--\eqref{eq:genproc_incidents}) and the code is specially to be noted. For the purpose of inference, Stan offers an automatic inference engine based on variational inference~\citep{kucukelbir2015} or sampling based algorithms~\citep{hoffman2014}. Results presented in the paper uses the latter. From a set of samples of $\big( \mathbf{w},\beta^i,\kappa^i_{X_1}, \kappa^i_{X_2}\big)$ obtained by HMC sampling, we relearn the hyperparameters ($\xi, k^i,\theta^i,\mathfrak{a}^i_{X_1},\mathfrak{b}^i_{X_1},\mathfrak{a}^i_{X_2},\mathfrak{b}^i_{X_2} $) in a maximum-likelihood setting using eqs.~\eqref{eq:prior_w}--\eqref{eq:prior_gamma2}. 

\subsubsection{Analytical method to obtain of $\pi_{post}\big(\mathbf{p}^i|\mathbf{ahl}^i,\mathbf{phl}^i\big)$} \label{subsubsec:Posterior_part2}
Here we describe an analytical method to update the posterior of Hurt level probabilities ($\textbf{p}^i$) given vectors of observed AHLs ($\textbf{ahl}^i$) and the corresponding PHLs  ($\textbf{phl}^i$) . 
\begin{subequations}
\begin{align}
\pi_{post}(\mathbf{p}^i|\mathbf{ahl}^i,\mathbf{phl}^i) & \propto \pi_{pr}(\mathbf{p}^i)\mathcal{L}(\mathbf{ahl}^i)\mathcal{L}(\mathbf{phl}^i) \\
&\propto \big[\pi_{pr}(\mathbf{p}^i)\mathcal{L}(\mathbf{ahl}^i) \big]\mathcal{L}(\mathbf{phl}^i)  \\
&\propto \big[ \pi_{post}(\mathbf{p}^i|\mathbf{ahl}^i)\big]\mathcal{L}(\mathbf{phl}^i) \label{eq:unnorm_posterior_p}
\end{align}
\end{subequations}
The prior on the $\textbf{p}^i$ is a Dirichlet distribution (refer to \equationref{eq:prior_p}) parametrized by a vector of pseudo-counts $\mathbf{\alpha}^i$. The bracketed term in   \equationref{eq:unnorm_posterior_p} admits an analytical form as a Dirichlet distribution (a standard result in Bayesian inference with a Dirichlet prior and a Multinomial likelihood),
\begin{equation} \label{eq:post_p_given_AHL} 
\pi_{post}(\mathbf{p}^i|\mathbf{ahl}^i) = \text{Dirichlet}(\mathbf{p}^i|\hat{\alpha}^i); \quad  \hat{\alpha}^i = \alpha^i + \mathbb{C}(\mathbf{ahl}^i)
\end{equation}
Here, the operator $\mathbb{C}(\cdot)$ denotes a counting operation, returning a vector of the number of time each AHL value is observed. This closed-form posterior acts as prior that now needs to be updated based on the PHL values. Note that PHL values are conditional on the corresponding AHL values (refer to \equationref{eq:genproc_PHL}), hence updating the hyper parameters ($\hat{\alpha}^i$) of the posterior is not that straightforward. We propose an update scheme that ensures that the following two characteristics are preserved.
\begin{itemize}
\item A PHL value informs the tail-end of the Hurt level distribution (the portion corresponding to Hurt levels $\geq$ AHL). In other words, a PHL value gives us insight about  the relative frequencies of Hurt levels  greater than the AHL, since it is a result of a post-incident scenario analysis that investigates the potential of a higher incident severity than the one encountered.
\item A PHL value does not impact the front-end of the Hurt level distribution (the portion corresponding to Hurt levels $<$ AHL) because by definition potentially more favorable outcomes of an incident than the observed AHL are not considered. 
\end{itemize}
Let $a\in\textbf{ahl}^i$, $p\in\textbf{phl}^i$, such that $p \geq a$, be the observed AHL and PHL values of a particular incident.  The hyper-parameters can be updated as follows:
\begin{subequations}
\begin{align}
\hat{\mathbf{\alpha}}^i[j] &= \hat{\mathbf{\alpha}}^i[j], \forall j < a  \label{eq:alpha_update_j_less_a}\\
\hat{\mathbf{\alpha}}^i[j] &= \big(\rho/(\rho+1)\big)  \times \hat{\mathbf{\alpha}}^i[j], \forall j \geq a, j \neq p \label{eq:alpha_update_j_more_a} \\
\hat{\mathbf{\alpha}}^i[j] &=   \big(\rho/(\rho+1)\big) \times (\hat{\mathbf{\alpha}}^i[j]+1),  j = p, \text{where}, \label{eq:alpha_update_j_equal_a} \\
\rho &=  \sum_{j \geq a} \hat{\mathbf{\alpha}}^i[j] 
\end{align}
\end{subequations}
Careful examination will reveal that \equationref{eq:alpha_update_j_less_a} and \equationref{eq:alpha_update_j_equal_a} embody the two aforementioned characteristics of the Hurt level distribution update given a PHL. Note that the cumulative pseudo-counts ($\sum_j \hat{\mathbf{\alpha}}^i[j]$) remains unchanged, as a result of which the marginal distributions of Hurt levels lower than $a$ remain unchanged. The final posterior distribution remains a Dirichlet distribution i.e. $ \pi_{post}(\mathbf{p}^i|\mathbf{ahl}^i,\mathbf{phl}^i) = \text{Dirichlet}(\mathbf{p}^i|\hat{\alpha}^i)$, where $\hat{\mathbf{\alpha}}^i$ is obtained via the applications of \equationref{eq:post_p_given_AHL} followed by eqs.~\eqref{eq:alpha_update_j_less_a}--\eqref{eq:alpha_update_j_equal_a}.

\subsection{Risk assessment and mitigation}\label{subsec:Optimization}

Given the state of vulnerabilities (in the form of posterior distributions) at any point of time, our goal is to quantify risk posed by the vulnerabilities.  Toward the end of \secref{sec:risk} we put forth the expected loss,  $s^i = \mathbb{E}(\mathbf{c}^T\textbf{ahl}^i)$, as one of the risk measures. Since, $\mathbf{c}$ is a vector comprising pre-specified loss values for all Hurt level, this expectation can be written equivalently as $\sum_i\mathbf{c}^T\mathbb{E}(\textbf{ahl}^i)$), where $\mathbb{E}$ is expectation operator with respect to the random variable $\mathbf{ahl}^i$. With certain assumptions, we  can obtain this expectation analytically as follows. From the proposed generative model (\equationref{eq:genproc_incidents} and \equationref{eq:genproc_AHL}) we can write, for the $i^{th}$ vulnerability,  the joint distribution of the number of incidents and the associated Hurt level count vector as follows.  
\begin{align}
p(n^i_\incidents , \mathbf{ahl}^i)  =  \text{Poisson}(n^i_\incidents | \kappa^i\beta^i) \ \text{Multinomial}(\mathbf{ahl}^i|n^i_\incidents,\mathbf{p}^i)
\end{align}
The above joint distribution can be expanded to include the parameters $\kappa^i, \beta^i, \mathbf{p}^i$ i.e. 
 \begin{align}\label{eq:joint_dist_all}
p(n^i_\incidents , \mathbf{ahl}^i,  \kappa^i\beta^i, \mathbf{p}^i)  =  \pi_{post}(\kappa^i\beta^i) \ \pi_{post}(\mathbf{p}^i) \  \text{Poisson}(n^i_\incidents | \kappa^i\beta^i) \ \text{Multinomial}(\mathbf{ahl}^i|n^i_\incidents,\mathbf{p}^i),
\end{align}
where $\pi_{post}(\cdot)$  represent the posterior distributions that are being updated at every time step (refer to \secref{subsec:Estimation}). The posterior of $\mathbf{p}^i$ is a Dirichlet distribution, with the concentration parameter $\mathbf{\alpha}^i$, i.e. $\pi_{post}(\mathbf{p}^i)=\text{Dirichlet}(\mathbf{p}^i|\mathbf{\alpha}^i)$, and is updated analytically as described in \secref{subsubsec:Posterior_part2}. On the other hand, the parameters $\kappa^i$ and $\beta^i$ are being sampled via the HMC algorithm (\secref{subsubsec:Posterior_part1}). We assume that the product of the later two parameters has a Gamma distribution, i.e. $\pi_{post}(\kappa^i\beta^i)=\text{Gamma}(\kappa^i\beta^i;\bar{k}^i,\bar{\theta}^i)$, with hyperparameters $\bar{k}^i$ and $\bar{\theta}^i$, which can be learned from the samples in a maximum-likelihood setting. This assumption allows us to obtain the expectation of $\mathbf{ahl}^i$ analytically as follows.
\begin{align}
\mathbb{E}(\textbf{ahl}^i) = \bar{k}^i\bar{\theta}^i\bigg(\frac{\mathbf{\alpha}^i}{\sum{\mathbf{\alpha}^i}}\bigg)
\end{align}
With this form of expectation, the expression of the risk measure (expected loss), for the $i^{th}$ vulnerability, can be written in a straightforward manner. This risk measure can be valuable to leadership when making safety related decisions, for instance, prioritizing allocation of risk-mitigation resources.  
\begin{align}\label{eq:expected_loss}
\mathbf{c}^T\mathbb{E}(\textbf{ahl}^i) = \bar{k}^i\bar{\theta}^i\bigg(\frac{\mathbf{c}^T\mathbf{\alpha}^i}{\sum{\mathbf{\alpha}^i}}\bigg)
\end{align}

In the remainder of this section we demonstrate a mathematical framework for automating such risk mitigation decisions. In the previous sections, we discussed how proactive data (safety observations) can be immensely useful to understand the overall safety landscape. However, it is not always clear how to efficiently use these limited observation resources (in the form safety inspectors). Assuming we can only make a finite number of observations everyday, we seek an optimal proportion of observations that should be allocated to each vulnerability such that an overall risk of future loss is minimized.  Mathematically, we can cast this resource allocation problem as an optimization problem, which when minimized (or maximized) yields an optimal proportions of resources to be allocated to each vulnerability. In the ensuing discussion we lay out the details of our proposed resource allocation method. Previously, we derived an expression for daily expected loss given by \equationref{eq:expected_loss}. It is this loss that can minimized for resource allocation, and hence forms the foundation the proposed resource allocation method. It is logical to assume that intervention of any form, for instance allocating certain number of observation resources, should transform the distribution $p(\mathbf{ahl}^i)$. Let $r^i$ represent the proportion of the resources associated to the $i^{th}$ vulnerability, we intend to define a conditional distribution   $p(\mathbf{ahl}^i|r^i)$. We posit that the influence of such intervention manifest through the distribution of the product $\kappa^i\beta^i$. As  a result, the joint distribution in  \equationref{eq:joint_dist_all} is now conditioned on $r^i$ as follows,
\begin{align}\label{eq:joint_dist_intervention}
p(n^i_\incidents , \mathbf{ahl}^i,  \kappa^i\beta^i, \mathbf{p}^i \ | \ r^i)  =  \pi_{post}(\kappa^i\beta^i  |  r^i) \ \pi_{post}(\mathbf{p}^i) \  \text{Poisson}(n^i_\incidents | \kappa^i\beta^i) \ \text{Multinomial}(\mathbf{ahl}^i|n^i_\incidents,\mathbf{p}^i).
\end{align}
The \equationref{eq:joint_dist_intervention} is identical to \equationref{eq:joint_dist_all} except that distribution of $\kappa^i\beta^i$ is now conditioned on the proportion of allocated resources $r^i$ . Recall from \secref{subsec:Model} that the product $\kappa^i\beta^i$ influences the incidence occurrence rate for the $i^{th}$ vulnerability (see  \equationref{eq:genproc_incidents}). Therefore, the impact of any intervention in terms of the reduction in incidence occurrence rate can be modeled by changing the distribution of $\kappa^i\beta^i$. As just discussed, the product of $\kappa^i\beta^i$ is characterized by a Gamma distribution with $\bar{k}^i \text{ and } \bar{\theta}^i$ as the hyperparameters .i.e. $\pi_{post}(\kappa^i\beta^i)=\text{Gamma}(\kappa^i\beta^i|\bar{k}^i,\bar{\theta}^i)$. We modify this distribution, by making it a function of resource proportions $r^i$, as follows
\begin{subequations}
\begin{align}
 \pi_{post}(\kappa^i\beta^i|r^i) &=\text{Gamma}\big(\kappa^i\beta^i \ | \ (1-h(r^i))^2\bar{k}^i,(1-h(r^i))^{-1}\bar{\theta}^i\big)  \label{eq:beta_conditional}\\
\text{where},\ & h(r^i) = \Big(1+\exp\big(-15(r^i-0.34)\big)\Big)^{-1}. \nonumber
\end{align}
\end{subequations}
The choice of this conditional distribution is rather ad~hoc but logical, and satisfies the following characteristics. The mean value of the product $\kappa^i\beta^i$ should decrease with the increase in resource proportion $r^i$ without necessarily changing its variance. Additionally, with half of the resources allocated to a vulnerability ($r^i=0.5$), the mean value should decrease by 90 percent. Similarly, with no resources allocated ($r^i=0$), the mean value more or less remains unchanged. A careful examination of the conditional distribution in  \equationref{eq:beta_conditional} would confirm it satisfies all of the aforementioned  characteristics. With this conditional distribution it is easy to verify that we can write the expectation of  $\mathbf{ahl}^i|r^i$ as follows
\begin{align}
\mathbb{E}(\textbf{ahl}^i|r^i) =  \big(1-h(r^i)\big) \bar{k}^i\bar{\theta}^i\bigg(\frac{\mathbf{\alpha}^i}{\sum{\mathbf{\alpha}^i}}\bigg)
\end{align} 
Finally, the resource allocation can be cast as the following optimization problem that aims to minimize the net-Expected loss. The objective function is differentiable can be minimized using any gradient-based optimization algorithm for instance quasi-Newton algorithms~\citep[Chapter~6]{nocedal2006}. This optimization problem is solved at every time interval (for example, everyday), with the most recently updated hyper-parameters viz.\ $\bar{k}^i, \bar{\theta}^i, \text{ and } \mathbf{\alpha}^i$, to yield optimal proportions of the resources to allocated to each vulnerability.
\begin{align}
\argmin_{r^1, r^2, \cdots r^n} &\sum_{i=1}^n   \big(1-h(r^i)\big) \bar{k}^i\bar{\theta}^i\bigg(\frac{\mathbf{c}^T\mathbf{\alpha}^i}{\sum{\mathbf{\alpha}^i}}\bigg) \\
\text{such that} & \  r^i>0 \text{ and } \sum_i r^i = 1 \nonumber
\end{align}

%% file: results.tex
\section{Case Studies}\label{sec:results}

In this section the effectiveness of the proposed method is demonstrated using two case studies. The first study is simulation-based, whereas the second uses real dataset from a large maintenance projects at a petrochemical plant. The proposed risk-based policy is compared to a \emph{random} and a \emph{heuristic-based} policy. The former, as the name suggests, is not informed by the safety data and assigns safety resources across vulnerabilities indiscriminately. This provides a lower-bound that any data-driven approach should surpass. While this is an uninformed approach, there is still a benefit because observation resources are being put to use. For this reason, the random policy is not the worst policy because it is unbiased and will direct some resources to vulnerabilities needing attention even if by chance. 

The heuristic-based policy, on the other hand, makes decisions in a data-driven fashion by computing a risk score of the $i^{th}$ vulnerability as a weighted sum of four aggregated quantitative indicators (between 0 and 1) of risk as shown in \equationref{eq:heuristic.score},
\begin{equation}
  s_i(t) = w_{obs}Q_{obs}^i(t) + w_{inc}Q_{inc}^i(t) + w_{ahl}Q_{ahl}^i(t) + w_{phl}Q_{phl}^i(t),
  \label{eq:heuristic.score}
\end{equation}
where $Q_{obs}^i(t)$ denotes, for vulnerability $i$, the ratio of the number of unsafe over all observations over a fixed period of time ending at time $t$ (e.g., 30~days). That is,
\begin{equation}
Q_{obs}^i(t) =  \frac{n^i_{a^-}(t)}{n^i_{a^-}(t)+n^i_{a^+}(t)}
\end{equation}
Note that the heuristic approach does not differentiate between observation types. $Q_{inc}$ reflects the number of incidents in the time period, including near-misses, according to
\begin{equation}
  Q_{inc}^i(t) = \begin{cases} 0.2n^i_e(t), & n^i_e(t)<5 \\ 1, & \text{otherwise} \end{cases}
\end{equation}
Then, $Q_{ahl}^i(t)$ and $Q_{phl}^i(t)$ account for the severity of incidents encountered. They correspond to the average Hurt-level severity (actual and potential, respectively) over the time period. The weights would be chosen by a user in a way that reflects how much they trust, or the relevance they put on observations versus incidents and number versus severity of incidents.
%Generally speaking, in this approach, near-miss and loss incidents tend to be weighted the highest.

\subsection{Validation in simulated environment}\label{subsec:sim_results}

It is easy to argue that when it comes to making safety related decisions, there could be a great risk in rolling out a new decision policy directly in the field. On one hand, any unanticipated shortcoming of a new policy may manifest in increased safety incidents. On the other hand, a subpar performance on a handful of occasions  may lead management to distrust a new policy, when in fact it could have significantly improved the safety performance in the long run.  This is where a simulation-based study becomes critical, by allowing a systematic vetting of a new policy under many different scenarios, while accounting for various real-world uncertainties. If carefully designed, any shortcoming/benefit of a new policy should become apparent in such a study, thereby informing the management before implementing a new decision policy. At the core of such a simulation study is a safety environment simulator that mimics the patterns in the occurrence of safety incidents typical to an industrial setting.  With built-in mechanisms to model external interventions, such a simulator provides an ideal and objective testbed to validate and compare different policies to make safety related decisions. One such simulation environment is proposed by \citet{paiva2021}, which is based on a detailed statistical model that captures various realistic and intuitive characteristics of the incidents and safety observations typical to construction and manufacturing industry. The readers should refer to the aforementioned reference for an in-depth look into the design of the simulator. Here we only provide the simulation settings under which the proposed method is tested and compared to the random and heuristic-based policies.

A few remarks on the use of a safety environment simulator to benchmark various decision policies.  First, the results from such comparison should be viewed, at best, in a qualitative sense. Any quantitative conclusion drawn, runs the risk of being incorrect. This is because the simulated environment is, almost surely, not going to be a true reflection of the reality. For instance, the simulator in the ensuing discussion comprises 7 vulnerabilities, while in reality there may be many more (possibly hundreds) of the same. Therefore, if one decision policy (to allocate observation resources) shows a 20\% reduction (with respect to the baseline) in a risk metric (such as \emph{expected-loss}, c.f. ~\secref{sec:risk}), one shouldn't expect the same level of reduction in reality. The simulation study only establishes relative trends one may expect to see if different decision policies are rolled out in the field. It is extremely hard to quantify the actual extent of the impact until after the roll out of a policy. Second, the proposed statistical model  that infers safety risks from both proactive and reactive safety data is markedly different from the model used to generate the safety data in the simulator. The latter is much more detailed and assumes comprehensive knowledge of the work environment such as the task and the incident rates of various vulnerabilities, biases in different observation types, the dynamics of the safety environment's evolution over time etc. The former, on the other hand, is a much simpler model that is postulated such that its free parameters can be inferred from scarce safety data. In other words, the model used to simulate the safety data is distinctly \textemdash and rightly so \textemdash different from the model used to assess safety risks and to subsequently allocate observation resources.

Consider a work environment with 7 vulnerabilities and 3 observation types.  Each of the seven vulnerabilities have distinct characteristics such as the number of tasks being carried out under them, propensity of safety incidents and their severity, and responsiveness to external intervention. These characteristics are controlled via certain parameters, which are specified in Table~\ref{tab:sim_setting_vuln}. Note that here we only specify the simulations parameters  for completeness, the detailed description of the simulated environment can be found in \citet{paiva2021}. With these parameters, each simulation is run for 365 days, during which the internal state of the safety environment (characterized by these parameters) is maintained continuously. Similarly, the three observations types WSO, BPO and SAO (refer to \secref{sec:observations} for the description of observation types), are characterized by certain parameters listed in Table~\ref{tab:sim_setting_obs}. At each time step (which is 1 day), a fixed number of observations (of each type) are made for every vulnerability. The maximum allowable observations per day are fixed for each observation type (see Table~\ref{tab:sim_setting_obs}). The actual allocation of observations to each vulnerability is determined by the policy under consideration.  The \emph{random} policy allocates resources in a completely random fashion, while the \emph{heuristic-based} and \emph{risk-based} policies take the historical data of safety incidents/observation in consideration for resource allocation. Anytime a negative observation is recorded for any vulnerability, the rate of occurrence of an incident corresponding to that vulnerability drops marginally in a pre-determined fashion (for an explanation, refer to Figure 3 in \citet{paiva2021}).  Logically, a good policy would allocate observation resources where the need is the most (i.e. to the vulnerabilities with high propensity of incidents or their severity or both) based on its assessment of the environment from the historical data.

\begin{table}%[tbhp]
\caption{A list of parameters of a simulated safety environment with 7 vulnerabilities. These parameter values fully specify the safety environment proposed in \citet{paiva2021}.}
\label{tab:sim_setting_vuln}
  \begin{tabular}{c*{11}{l}}
    \toprule
    & \multicolumn{11}{c}{Simulator Parameters} \\ \cline{2-12}
%\begin{tabular}{| c | l | l | l |  l | l | l | l | l | l | l | l |}
%\hline
%vulnerability\textbackslash{}parameters & \multicolumn{1}{l|}{$\lambda_*$} & \multicolumn{1}{l|}{$\xi_{base}$} & \multicolumn{1}{l|}{$\theta[0]$} & \multicolumn{1}{l|}{$k$} & \multicolumn{1}{l|}{$\alpha$} & \multicolumn{1}{l|}{$p_{hl0}$} & \multicolumn{1}{l|}{$p_{hl1}$} & \multicolumn{1}{l|}{$p_{hl2}$} & \multicolumn{1}{l|}{$p_{hl3}$} & \multicolumn{1}{l|}{$p_{hl4}$} & \multicolumn{1}{l|}{$p_{hl5}$} \\ \hline
    Vulnerability & $\lambda_*\ $ & $\xi_{base}$ & $\theta[0]$ & $k$ & $\alpha$
      & $p_{0}$ & $p_{1}$ & $p_{2}$ & $p_{3}$ & $p_{4}$ & $p_{5}$ \\
    \midrule
A                                      & 17                               & 0.55                             & 0.31                             & 0.97                     & 0.04                          & 0.5                            & 0.35                           & 0.13                           & 0.02                           & 0                              & 0                              \\ %\cline{1-1}
B                                      & 13                               & 0.25                             & 0.88                             & 0.99                     & 0.005                         & 0.6                            & 0.11                           & 0.11                           & 0.16                           & 0.02                           & 0                              \\ %\cline{1-1}
C                                      & 22                               & 0.4                              & 0.53                             & 0.95                     & 0.01                          & 0.3                            & 0.05                           & 0.35                           & 0.28                           & 0.02                           & 0                              \\ 
%\cline{1-1}
D                                      & 12                               & 0.1                              & 0.45                             & 0.98                     & 0.005                         & 0.2                            & 0.3                            & 0.25                           & 0.18                           & 0.04                           & 0.03                           \\ %\cline{1-1}
E                                      & 15                               & 0.05                             & 0.18                             & 0.99                     & 0.01                          & 0.2                            & 0.16                           & 0.16                           & 0.16                           & 0.16                           & 0.16                           \\ %\cline{1-1}
F                                      & 5                                & 0.45                             & 0.78                             & 0.97                     & 0.02                          & 0.4                            & 0.03                           & 0.08                           & 0.18                           & 0.26                           & 0.05                           \\ 
%\cline{1-1}
G                                      & 22                               & 0.3                              & 0.35                             & 0.99                     & 0.005                         & 0.65                           & 0.15                           & 0.08                           & 0.06                           & 0.04                           & 0.02                           \\
    \bottomrule
\end{tabular}
\end{table}

\begin{table}%[tbhp]
\centering
\caption{Simulation parameters with respect to each observation type. The parameters $m$ and $\delta_{a-}$ denote the number of observations made per day and the relative reduction in incident rate of a vulnerability when a negative observation is made, respectively. The parameters $\eta_{a-}$ and $\eta_{a+}$ model the biases in the observation process. In the current setting, WSO and BPO observation types are biased (the former is biased towards producing more negative observations and the latter vice versa), while SAO is unbiased.}
\label{tab:sim_setting_obs}
  \begin{tabular}{ccccc}
    \toprule
    Observation Types & $m$ & $\delta_{a-}$ & $\eta_{a-}$ & $\eta_{a+}$ \\
    \midrule
    WSO & 2 & 0.03 & 100 & 150 \\
    SAO & 2 & 0.03 & 100 & 100 \\
    BPO & 1 & 0.03 & 120 & 100 \\
    \bottomrule
  \end{tabular}
\end{table}
 
During a simulation run, with each of the three policies, various metrics that reflect the risk associated with each vulnerability were recorded on daily basis. Two such metrics are the \emph{tail probability} and the \emph{expected-loss}. The former denotes the probability of an incident with AHL $\geq 4$ to occur every day. The latter quantifies the average daily loss that is likely to incur due to safety incidents. We have alluded to the expected-loss risk metric in ~\secref{sec:risk}, but for a detialed exposition on this subject, refer to Section 3.4 in \citet{paiva2021}. 

\begin{figure}%[!tbp]
 %\centering
 \includegraphics[width=\textwidth]{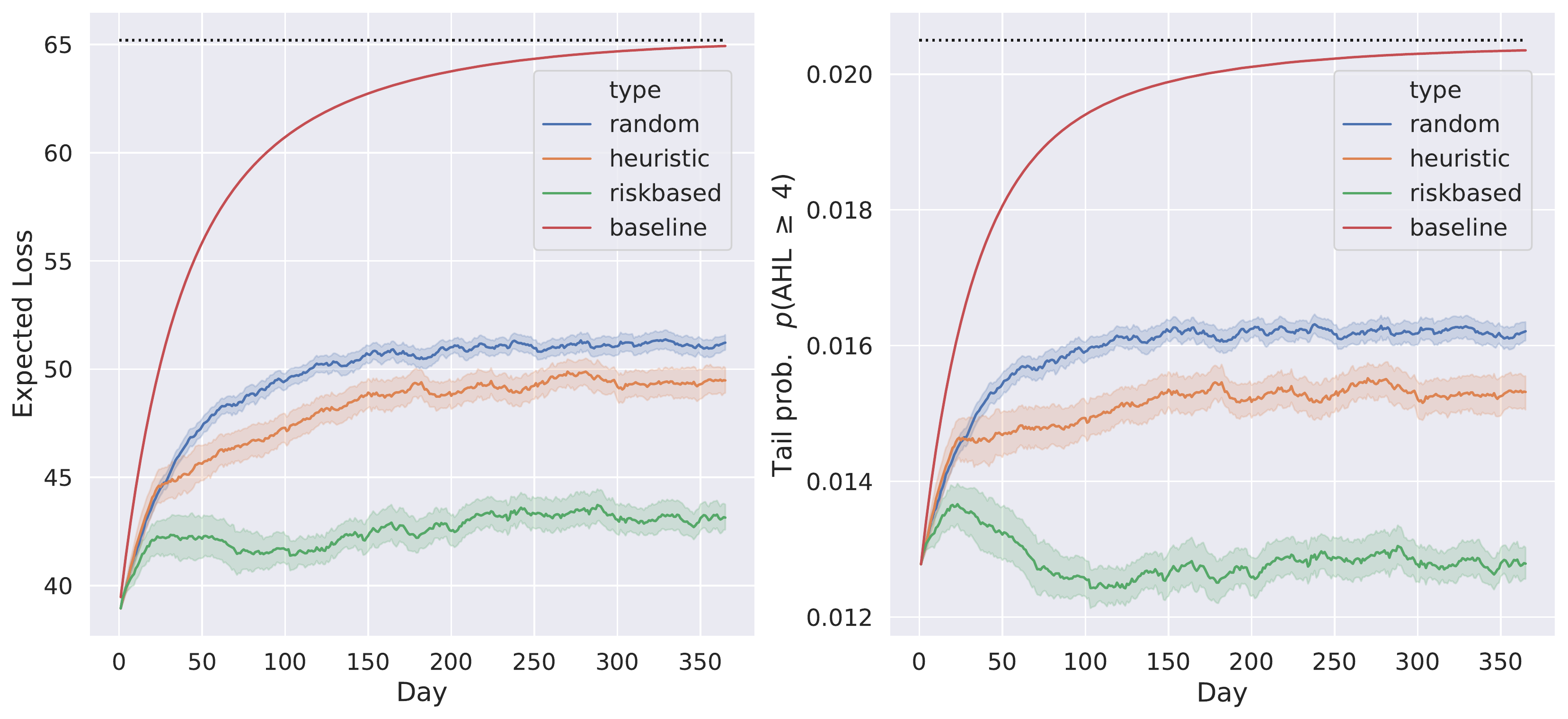}
  \caption{Comparison of the three policies for allocating observation resources on two risk metrics viz.  \emph{expected-loss} and \emph{tail-probability}. The baseline correspond to the \emph{no-observation} case i.e. when zero resources are  allocated for safety observations, due to which the environment health deteriorates at a natural rate as governed by the simulation. The plots are obtained by running simulations 100 times from the same initial state. The solid lines and the bands denote the mean value and the 95$\%$ confidence interval on the estimated mean, respectively. Note that the baseline plots are deterministic since there is no feedback (since no observations), in which case the expected-loss and tail-probability evolve deterministically. Also, the dotted lines represent the limiting (or the worst-case) values of risk metrics for the baseline case.}
 \label{fig:tail_prob_and_exp_loss}
\end{figure}

Figure~\ref{fig:tail_prob_and_exp_loss} compares the performances of various policies in the aforementioned simulated environment. The baseline curves correspond to the case when no safety observations are made to monitor \textemdash and subsequently discourage\textemdash the unsafe practices and behaviors which may exists in the field. This can be considered as the worst case scenario, because without any monitoring (and feedback) the work environment's safety continuously deteriorates at some natural rate (determined by the simulation parameters). Additionally, in the baseline scenario the two risk metrics evolve deterministically since in absence of safety observations the environment gets no feedback and thus its state evolves predictably. On the other hand, when safety observations are made (based on any policy), the resulting feedback changes the state of the environment, thus adding stochasticity to the two risk metrics. To account for this stochasticity we ran 100 independent simulations with the same initial state for all the policies in consideration. The time series plots in Figure ~\ref{fig:tail_prob_and_exp_loss} show the mean values of the two risk metrics and the 95$\%$ confidence interval around them. Evidently, the random policy and the risk-based policy fare the worst and the best, respectively, with the heuristic-based policy somewhere in between. This observation is  consistent with the intuition because the resource-allocation decisions made by the random policy are not data-driven unlike the other two. Before we further analyze the relative performances of different policies a quick remark on the chosen initial state of the simulated safety environment. Note that the safety environment starts well with low values of tail probability and expected-loss metrics. Thereafter, regardless of the policy in consideration, the safety performance degrades simply because the available observation resources (refer to Table \ref{tab:sim_setting_obs}) are not enough to offset the natural deterioration built in the simulated safety environment. Nevertheless, risk-based policy makes better use of the limited resources over time, which progressively drives the safety performance up, as evident in the plots in Figure \ref{fig:tail_prob_and_exp_loss}.  

%\begin{figure} %[!tbp]
%  \centering
%  \subfloat[Tail probability]{\includegraphics[width=0.50\textwidth]{tail_prob_various_policies_sim1_seed40}\label{fig:f1}}
%  \hfill
%  \subfloat[Expected loss]{\includegraphics[width=0.48\textwidth]{exp_loss_various_policies_sim1_seed40}\label{fig:f2}}
%  \caption{Comparison of the three policies for allocating observation resources on two health measures viz. \emph{tail probability} and \emph{expected loss}. The baseline correspond to the \emph{no-observation} case i.e. when zero resources are allocated for safety observations, due to which the environment health deteriorates at a natural rate as governed by the simulation. The plots are obtained by running simulations 100 times from the same initial state. The solid lines and the bands denote the mean value and the 95$\%$ confidence interval on the estimated mean, respectively. Note that the baseline plots are deterministic since there is no feedback (since no observations), in which case the expected loss and tail probabilities evolve deterministically.}
%  \label{fig:tail_prob_and_exp_loss}
%\end{figure}

\begin{figure} %[!tbp]
  \includegraphics[width=1.1\textwidth]{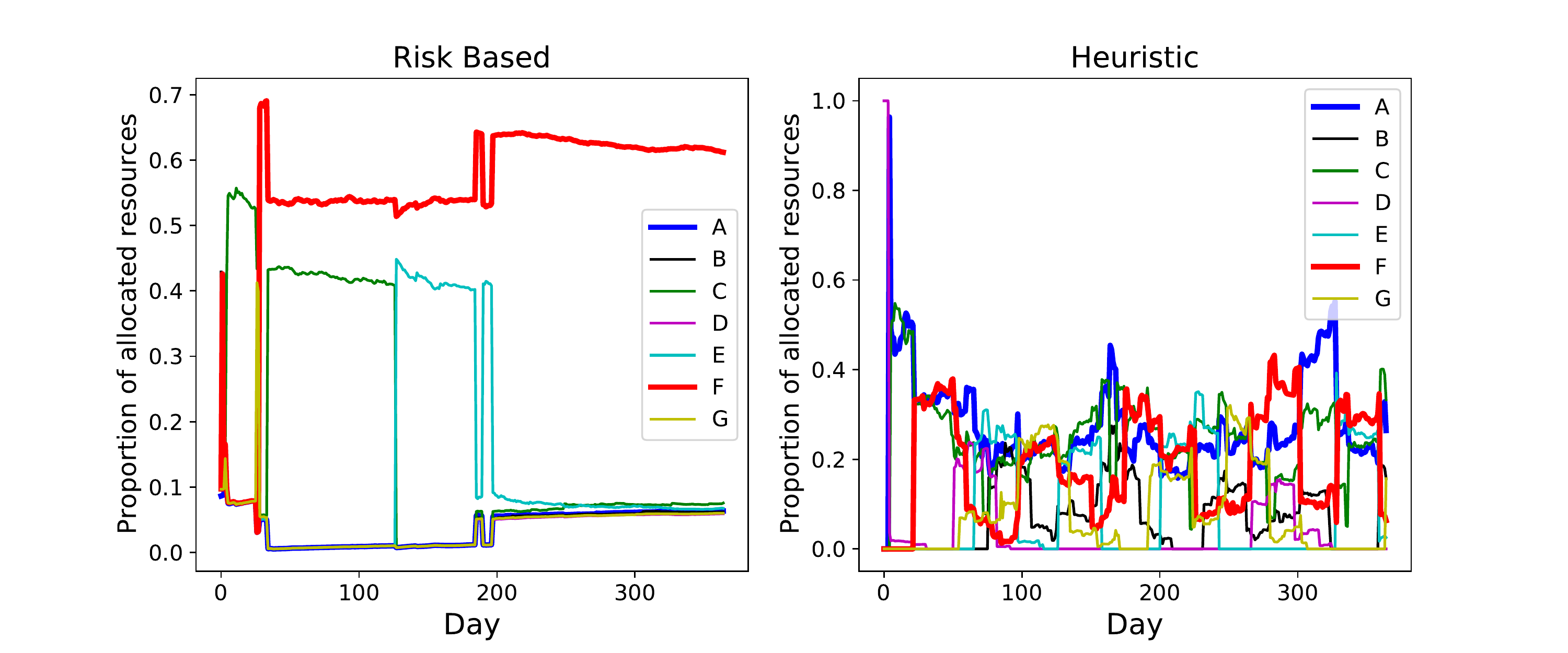}
  \caption{The proportion of observation resources allocated to every vulnerability by the risk-based (left) and heuristic(right) policies. This plots corresponds to one simulation run. To highlight the difference between the two policies, the time series of resource proportions are accentuated for vulnerability A (blue) and F (red). The strength of risk-based policy is clearly evident by the manner in which it learns the higher risk associated with vulnerability F, thereby allocating more resources to it. The heuristic policy lacks this learning capability and is merely reactive to the safety data from recent history.}
  \label{fig:vuln_scores}	
\end{figure}

To further understand the superior performance of the risk-based policy compared to the heuristic policy, we would like to draw a reader's attention to figure \ref{fig:vuln_scores}. The plots in this figure show the fraction of resource allocated (to each of the 7 vulnerability) by the two policies over the duration of one simulation. For the ease of readability we highlight the fraction of resources allocated to two vulnerabilities viz.\ A (in blue) and F (in red). These two vulnerabilities are starkly different in the sense that vulnerability A represents a safety area under which numerous observations and incidents are reported (for instance \emph{hand safety}) , but severity of incidents is rather low (e.g. incidents like paper-cuts). On the other hand, vulnerability~F represents an area that is infrequently encountered but is more prone to human errors and also results in high severity incidents (e.g. \emph{crane lift}). The two policies allocate the resources very differently. The risk-based policy learns (from the historical data) the fact that vulnerability F has a higher propensity of human error and high severity incidents, and thus persistently allocates more resources to it. Similarly, vulnerability~A is allocated a small fraction of resources since it is relatively inconsequential. The heuristic policy is unable to distinguish the two vulnerabilities, and in fact allocates more resources to vulnerability~A than F. This is primarily because of the ad-hoc nature of this policy, which although is data-driven (by accounting for all the factors such as negative observations, incident counts, AHL and PHLs), does not explicitly model the inherent risk of each vulnerability. Additionally, the heuristic policy has a limited memory (defined by a window size) which makes it oblivious to true risks if the data in the recent history happens to show no red flags.

\begin{figure} %[!tbp]
  \centering
  \includegraphics[width=0.6\textwidth]{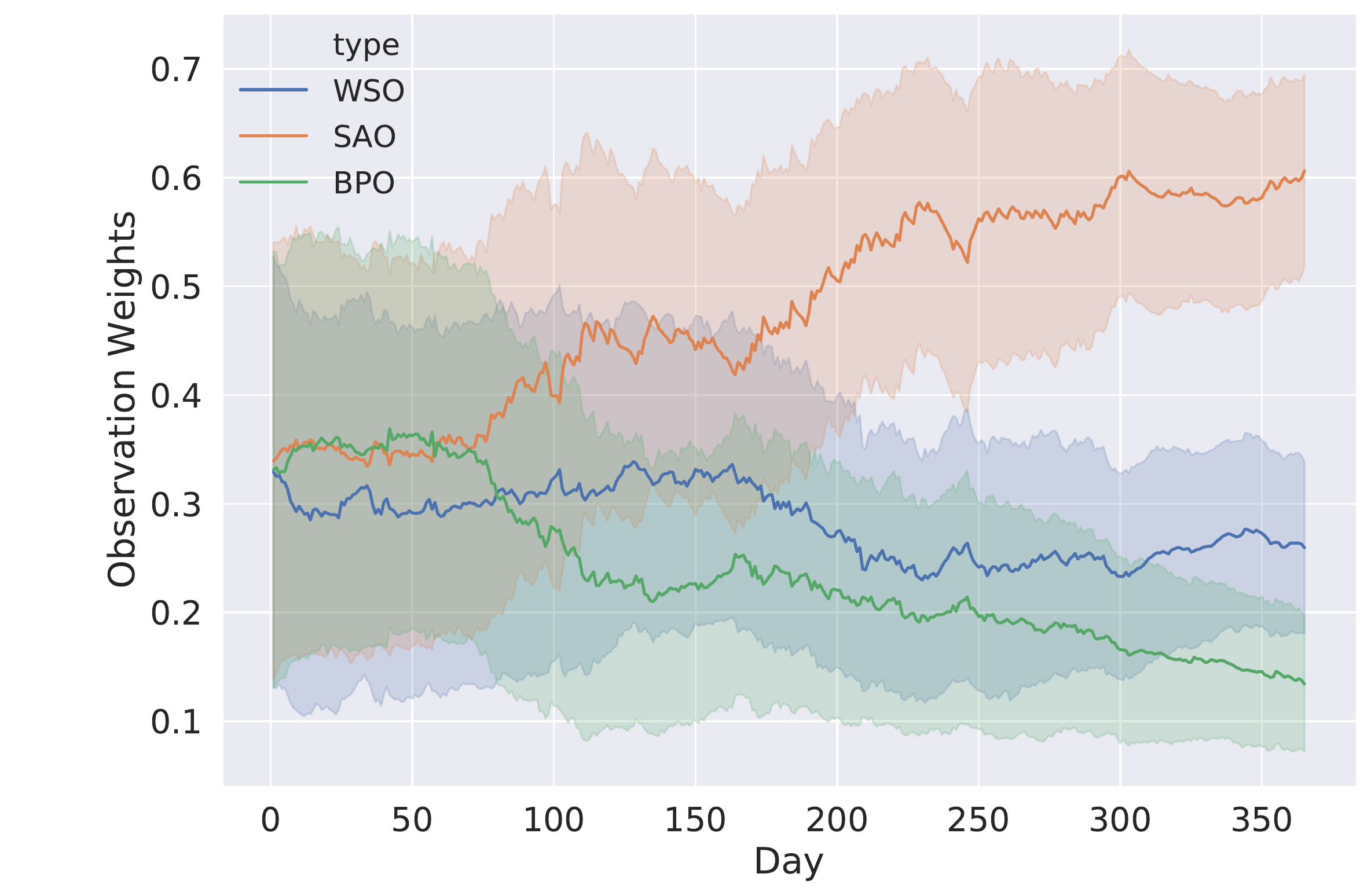}
  \caption{Time evolution of weights of the three observation types as inferred by the proposed risk-based approach. Starting from an initial state, where all observation are equi-weighted, the approach continuously updates the weights of all observation types as the new safety data arrives. The line-plots show the mean value and $\pm1$ standard deviation uncertainty band. As more data is observed the uncertainty of the weights gradually diminishes as expected.}
  \label{fig:obs_weights}	
\end{figure}

A useful byproduct of the risk-based approach is that it internally weighs different observation types and infers these weights from historical data in an online fashion (see \equationref{eq:genproc_incidents} and exposition in section \ref{subsec:Estimation} for details). Figure \ref{fig:obs_weights} plots the time evolution of these weights for one run of the simulation. Starting from equal weights with high uncertainty, the approach (after ~100 days) deemed SAO observations to have a higher weight in terms of their ability to explain the occurrence of incidents. Additionally, the uncertainty around the weights reduces with the passage of time as more safety data is observed. These results are inline (qualitatively) with the simulation setting, where SAO was the only unbiased observation type (refer to Table~\ref{tab:sim_setting_obs}). Such additional piece of information can be crucial for the management to assess the efficacy of different observation processes in place.

\begin{figure} %[!tbp]
  \centering
  \includegraphics[width=\textwidth]{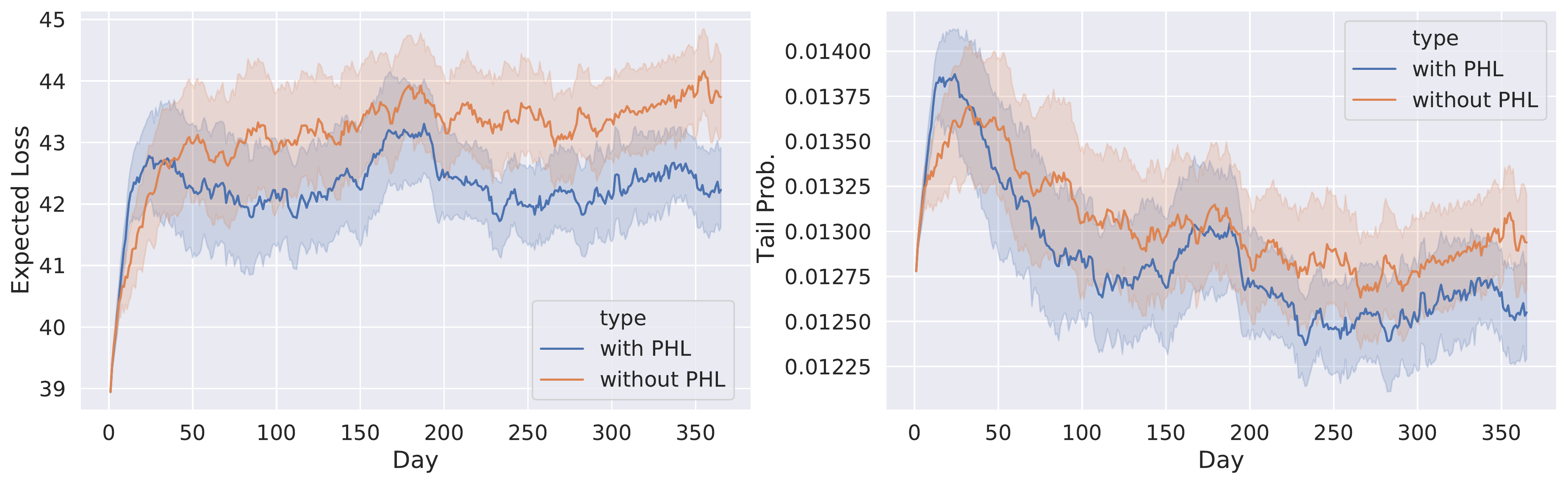}
  \caption{Time evolution of the two health measures viz. \emph{expected loss} (left) and \emph{tail probability}, under the risk-based allocation approach. Two scenarios are considered i.e. when the vulnerabilities' states are updated with and without the PHL information. One hundred simulations are used to generate the line-plots of the mean values of the two health measures. The band shows the standard error in the mean estimates. The simulation setting is identical to the one used in \figref{fig:tail_prob_and_exp_loss}.}
  \label{fig:PHL_impact}	
\end{figure}

Finally, one of the strengths offered by the proposed approach is to assimilate information from PHLs to assess the risks posed by vulnerabilities. A PHL is a product of systematic critiquing of the work environment by considering the worst case outcome of an incident. It is widely acknowledged that such information can help eliminate blind spots by being a leading indicator of incidents in future. However, including information from PHLs during risk assessment is tricky because there is no prescribed way to weigh this information along with other sources of safety information such as AHLs, incident counts and safety observations. In section \ref{subsubsec:Posterior_part2} we detailed a systematic and logical way to incorporate PHL information to update the state of a vulnerability. The impact of having this information is quantified in our simulation study and is shown in figure \ref{fig:PHL_impact}. For the same test-bed used to generate figure \ref{fig:tail_prob_and_exp_loss}, we run the risk-based allocation with and without incorporating the PHL information. It is evident that including PHLs improves the performance of risk-based resource allocation in a statistically significant manner. One may argue that the improvement is marginal. A counter-argument to that is even a small improvement in the area of industrial safety is highly desired, especially if that is coming with little or no additional cost. Additionally, as previously mentioned, it is not recommended to draw quantitative conclusions from such simulation studies. This result merely establishes, in a qualitative sense, that including PHL information in the risk-based approach improves the safety performance. 

\subsection{Real industrial data}\label{subsec:real_results}

\begin{figure}[tbp]
  \centering
  \includegraphics[width=\textwidth]{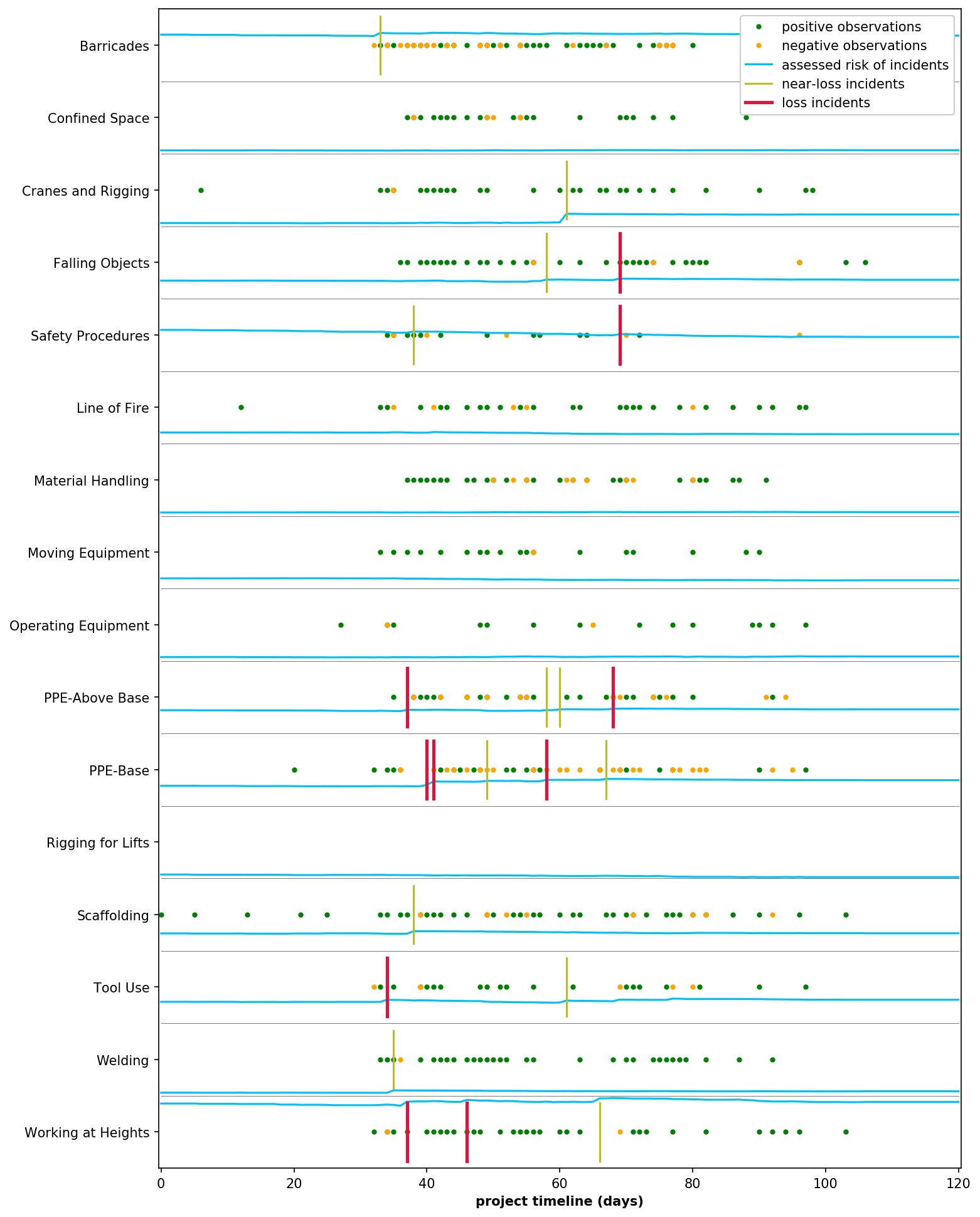}
  \caption{Real data and expected-loss risk metric over time for Project~A. The different vulnerabilities are shown in the vertical axis. For each vulnerability, dots are shown for the days where observations were collected with the color of dot denoting whether the observation identified safe aspects (in green) or unsafe aspects (in orange). Near-miss incidents are shown as thin green vertical bars and incidents as thick red vertical bars. The blue line of each vulnerability denotes the expected-loss metric (\equationref{eq:expected_loss}) normalized between 0 and 1. Although the analysis comprises 60 vulnerabilities, only 16 of them are plotted for clarity.}
  \label{fig:real.data1:csr}
\end{figure}

\begin{figure}[tbp]
  \centering
  \includegraphics[width=\textwidth]{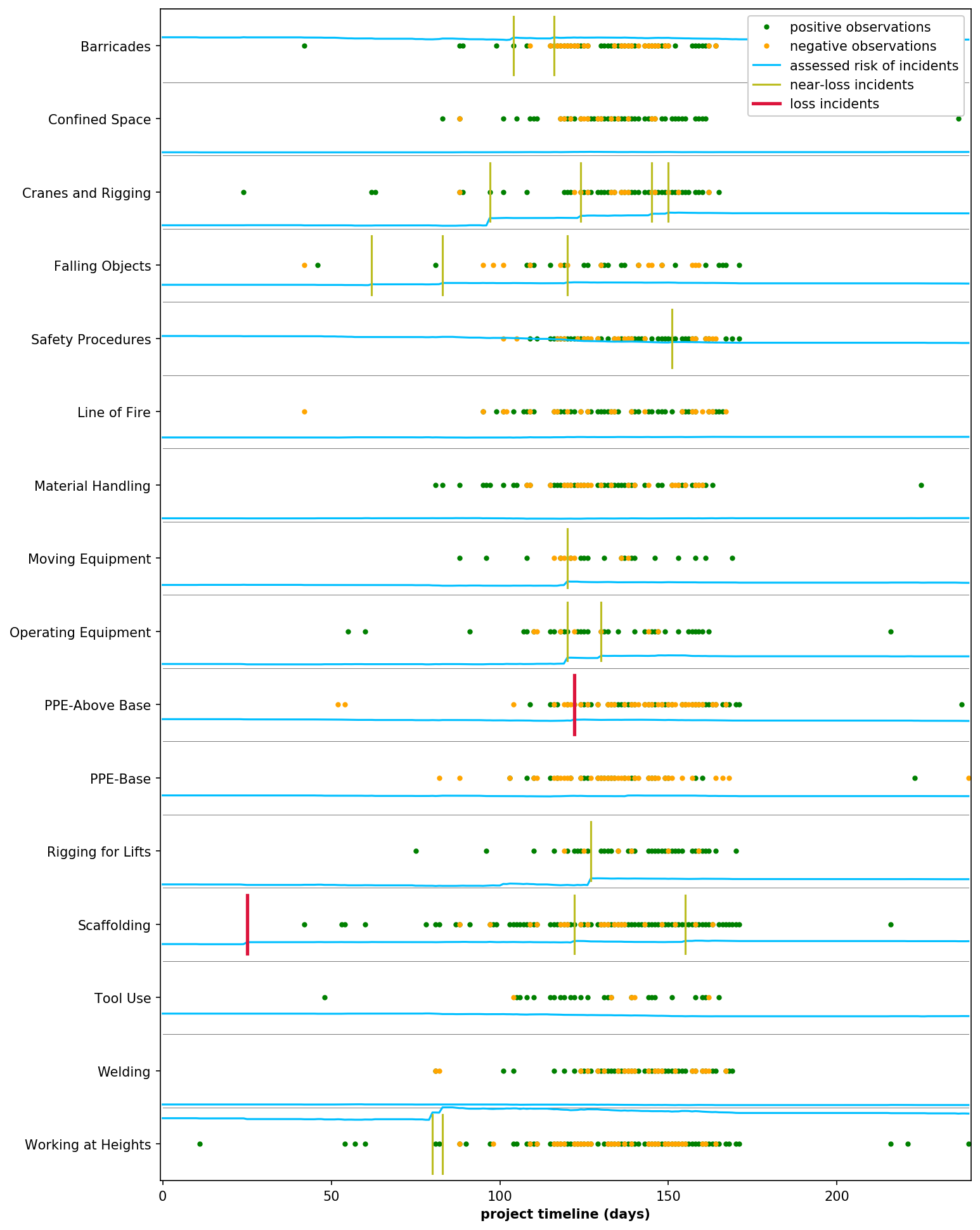}
  \caption{Real data and expected-loss risk metric over time for Project~B, as in \figref{fig:real.data1:csr}.}
  \label{fig:real.data2:csr}
\end{figure}

The proposed risk-based approach is now applied to a real-world scenario. Of course, in such scenarios the true underlying risks are not known, and thus the results cannot be evaluated quantitatively. Therefore, the results shown here are not be taken at face value as they may not represent the true relative risks of various workplace hazards. Such risks may greatly vary from one work-site to another, thereby necessitating some customization of the proposed method. Nonetheless, the purpose of the ensuing analysis is to show that the proposed approach systematically learns and adapts its assessments of safety risks from historical data, and the consistency and robustness of those assessments under a variety of conditions. To this end, we consider data from two large maintenance projects at a petrochemical manufacturing plant. The application in these case studies is meant to demonstrate how the presence of different vulnerabilities (60 in this study) and the number of observations and incidents influence the assessment of safety risks over time. The safety data of these maintenance projects, referred as Project~A and B, is depicted in \figsref{fig:real.data1:csr} and \ref{fig:real.data2:csr}. In these figures, and for each vulnerability, the days with safety observations or incidents are shown as dots and vertical bars, respectively. As can be observed, the data is quite sparse, with very few observations, especially during earlier and later periods of the project. Thanks to the efficacy of current efforts, the number of incidents is also small. There is also substantial variability in the amount of data collected between vulnerabilities over time. As previously mentioned, these are some intrinsic challenges that a data-driven method to assess safety risks must robustly handle. Also shown in \figsref{fig:real.data1:csr} and \ref{fig:real.data2:csr} is the expected-loss metric (refer to \equationref{eq:expected_loss} and \secref{subsec:Optimization} for more details) for each vulnerability shown by the blue line. The daily expected-loss was estimated in a causal manner, meaning that the safety data up until that day was used as input to the model. The hyper-parameters of the proposed hierarchical, probabilistic model were calibrated using data from two previous maintenance projects (not shown here), and the final calibrated values were then used as the prior values for obtaining the results shown. Since the hyper-parameters are adapted from the observed data at every time step (a day in this case), at the end of a project, they more accurately capture the characteristics of each vulnerability as well as any potential biases in the way the safety data is acquired in the work environment under consideration. A quick remark on the importance of appropriate initialization of the hyper-parameters.  Since these parameters quantitatively encode our prior beliefs about various safety risks, they can be initialized by a safety manager who is often equipped with a holistic knowledge of the work environment that may not be present in the limited, historical safety data. Therefore, like any other Bayesian method, a hallmark of our risk assessment approach is that it provides a great opportunity to combine subjective, yet substantive, information with the field data in a systematic fashion. 

The assessed risk quantified by the expected-loss metric clearly shows certain vulnerabilities that standout from the rest. For instance, \emph{Working at Heights} systematically has the highest risk throughout the duration of either project. That is a combination of the fact that incidents associated with this vulnerability have had higher severity, and that they continue to occur with relatively high likelihood. In contrast, the risk of PPE-related vulnerabilities (\emph{PPE-Base} and \emph{PPE-Above Base}) remains in the mid to low range, despite having the highest number of incidents but with much lower severity. This means that, although it is important to continue to reinforce PPE safe practices, in this work environment, there are other vulnerabilities of potentially higher concern. In both projects, the vulnerabilities with the second and third highest risks were \emph{Barricades} and \emph{Safety Procedures}. Although the two vulnerabilities have similar a~priori risk and had a similar number incidents, the observations pertaining to \emph{Barricades} frequently identify unsafe practices or behaviors. Thus, between them, the assessed risk of the former is marginally higher than the latter. 

To corroborate the consistency and robustness of the proposed method to the sparseness of safety data, consider again the two vulnerabilities \emph{Barricades} and \emph{Safety Procedures}. The estimated risk of the two remained comparable in both projects, and this occurs in spite of the number of observations for the latter being smaller than those for the former, especially in Project~A. Another example corresponds to the  assessed risk of \emph{Cranes and Rigging}, which stabilizes at comparable levels in both projects even though there were four near-misses in Project~B and only one in Project~A. In reality the true underlying risks of vulnerabilities shouldn't change drastically from one project to another unless the projects are of different nature, or the workforce has changed significantly, or the projects are far apart in time. The proposed method  seems to preserve this consistency of assessed risk across different projects despite seeing different amounts of data.  At the same time, if the projects do involve different work types, an ideal method should reflect that in its corresponding risk assessments. This can be noted in the assessed risks of the \emph{Moving Equipment} and \emph{Rigging for Lifts} vulnerabilities. In Project~A, the observations in these areas are quite sparse (even zero for the latter)  and there were no incidents. This could indicate that equipment handling was perhaps not a large aspect of that maintenance project. Therefore, the assessed risks of these vulnerabilities remained low throughout.  On the other hand, in Project~B (\figref{fig:real.data2:csr}), there is relatively larger number of observations and corresponding near-misses, thus the model revises the corresponding risk estimates accordingly.  

The  assessed risk curves in \figsref{fig:real.data1:csr} and \ref{fig:real.data2:csr} also highlight the model adaptation with different forms of safety data. If only observations (proactive data) are available, the risk estimates change over time albeit slowly because such information can be ``noisy'' (subject to human bias). However, once incidents (reactive data) are observed, they provide crucial and unbiased information, which is then used to revise several of the model's parameters and the subsequent estimated risks. This aspect can be easily seen with regard to the risk curves for the \emph{Cranes and Rigging} vulnerability, in either project. Note how the assessed risk in Project~B adjusts with every near-miss incidents, however the adjustments are increasingly smaller. This happens, at least in part, because the latter incidents are  better explained by the  updated model, and does not lead to significant changes in its parameters. In contrast, note how the assessed risk for the same vulnerability in Project~A (\figref{fig:real.data1:csr}) was adjusted abruptly after the only near-miss because unlike Project~B there was a large number of recent safe observations and yet the near-miss incident occurred. This data revealed an incongruence in the model's assessment, resulting in a sharp revision of the risk henceforth.

Finally, an important aspect of proposed method is that it allows adaptation in a dynamic environment, thereby enabling identification of evolving risks of various occupational hazards. This is exemplified by  the assessed risks of \emph{Falling Objects}, \emph{PPE-Base} and \emph{Working at Heights} in Project~A. In all these cases, the risk increases (sometimes marginally) as the incoming safety data comprises of actual/near-miss incidents and negative observations. In all these cases, there are subsequent incidents in future, which could indicate that the proposed method is indeed able to identify evolving risks in a dynamic work environment.

%% file: paper.bbl
\begin{thebibliography}{44}
\expandafter\ifx\csname natexlab\endcsname\relax\def\natexlab#1{#1}\fi
\expandafter\ifx\csname url\endcsname\relax
  \def\url#1{\texttt{#1}}\fi
\expandafter\ifx\csname urlprefix\endcsname\relax\def\urlprefix{URL }\fi

\bibitem[{Arslan et~al.(2020)Arslan, Cruz, and Ginhac}]{arslan2020identifying}
Arslan, M., Cruz, C., Ginhac, D., 2020. Identifying intrusions in dynamic
  environments using semantic trajectories and {BIM} for worker safety. In:
  Fourth International Congress on Information and Communication Technology.
  Springer, pp. 59--67.

\bibitem[{Awolusi et~al.(2018)Awolusi, Marks, and
  Hallowell}]{awolusi2018wearable}
Awolusi, I., Marks, E., Hallowell, M., 2018. Wearable technology for
  personalized construction safety monitoring and trending: Review of
  applicable devices. Automation in construction 85, 96--106.

\bibitem[{Baradan and Usmen(2006)}]{baradan2006}
Baradan, S., Usmen, M.~A., 2006. Comparative injury and fatality risk analysis
  of building trades. Journal of Construction Engineering and Management
  132~(5), 533--539.

\bibitem[{Carpenter et~al.(2017)Carpenter, Gelman, Hoffman, Lee, Goodrich,
  Betancourt, Brubaker, Guo, Li, and Riddell}]{carpenter2017}
Carpenter, B., Gelman, A., Hoffman, M.~D., Lee, D., Goodrich, B., Betancourt,
  M., Brubaker, M., Guo, J., Li, P., Riddell, A., 2017. Stan: A probabilistic
  programming language. Journal of Statistical Software, Articles 76~(1).

\bibitem[{Chae and Yoshida(2010)}]{chae2010application}
Chae, S., Yoshida, T., 2010. Application of {RFID} technology to prevention of
  collision accident with heavy equipment. Automation in construction 19~(3),
  368--374.

\bibitem[{Cheng et~al.(2013)Cheng, Yao, and Wu}]{cheng2013applying}
Cheng, C.-W., Yao, H.-Q., Wu, T.-C., 2013. Applying data mining techniques to
  analyze the causes of major occupational accidents in the petrochemical
  industry. Journal of Loss Prevention in the Process Industries 26~(6),
  1269--1278.

\bibitem[{Cornell(1968)}]{cornell1968}
Cornell, C.~A., 10 1968. Engineering seismic risk analysis. Bulletin of the
  Seismological Society of America 58~(5), 1583--1606.

\bibitem[{Curcuruto et~al.(2015)Curcuruto, Conchie, Mariani, and
  Violante}]{curcuruto2015role}
Curcuruto, M., Conchie, S.~M., Mariani, M., Violante, F., 2015. The role of
  prosocial and proactive safety behaviors in predicting safety performance.
  Safety science 80, 317--323.

\bibitem[{Daley et~al.(1988)Daley, Vere-Jones, and W}]{daley1988introduction}
Daley, D., Vere-Jones, D., W, V., 1988. An Introduction to the Theory of Point
  Processes. Probability and its applications. Springer-Verlag.

\bibitem[{Dekker(2016)}]{dekker2016drift}
Dekker, S., 2016. Drift into failure: From hunting broken components to
  understanding complex systems. CRC Press.

\bibitem[{Dekker and Pitzer(2016)}]{dekker2016examining}
Dekker, S., Pitzer, C., 2016. Examining the asymptote in safety progress: a
  literature review. International Journal of Occupational Safety and
  Ergonomics 22~(1), 57--65.

\bibitem[{Desvignes(2014)}]{desvignes2014}
Desvignes, M., 2014. Requisite empirical risk data for integration of safety
  with advanced technologies and intelligent systems. Ph.D. thesis, University
  of Colorado.

\bibitem[{Eggleston et~al.(2014)Eggleston, Hayes,
  et~al.}]{eggleston2014assessing}
Eggleston, V., Hayes, B., et~al., 2014. Assessing the actual and potential
  environmental and socioeconomic effects of incidents. In: SPE International
  Conference on Health, Safety, and Environment. Society of Petroleum
  Engineers.

\bibitem[{Esmaeili and Hallowell(2012)}]{esmaeili2012diffusion}
Esmaeili, B., Hallowell, M.~R., 2012. Diffusion of safety innovations in the
  construction industry. Journal of Construction Engineering and Management
  138~(8), 955--963.

\bibitem[{Etaje et~al.(2013)Etaje, Abdulkarim, Ibe,
  et~al.}]{etaje2013efficiency}
Etaje, D., Abdulkarim, M., Ibe, J., et~al., 2013. The efficiency of hurt based
  approach in improving personnel safety. In: SPE Nigeria Annual International
  Conference and Exhibition. Society of Petroleum Engineers.

\bibitem[{Fugas et~al.(2012)Fugas, Silva, and Meliá}]{fugas2012468}
Fugas, C.~S., Silva, S.~A., Meliá, J.~L., 2012. Another look at safety climate
  and safety behavior: Deepening the cognitive and social mediator mechanisms.
  Accident Analysis \& Prevention 45, 468 -- 477.

\bibitem[{Goh and Ubeynarayana(2017)}]{goh2017construction}
Goh, Y.~M., Ubeynarayana, C., 2017. Construction accident narrative
  classification: An evaluation of text mining techniques. Accident Analysis \&
  Prevention 108, 122--130.

\bibitem[{Grant et~al.(2018)Grant, Salmon, Stevens, Goode, and
  Read}]{grant2018back}
Grant, E., Salmon, P.~M., Stevens, N.~J., Goode, N., Read, G.~J., 2018. Back to
  the future: What do accident causation models tell us about accident
  prediction? Safety Science 104, 99--109.

\bibitem[{Haas et~al.(2020)Haas, Demich, and McGuire}]{haas2020learning}
Haas, E.~J., Demich, B., McGuire, J., 2020. Learning from workers near-miss
  reports to improve organizational management. Mining, Metallurgy \&
  Exploration, 1--13.

\bibitem[{Hallowell and Gambatese(2010)}]{hallowell2010}
Hallowell, M.~R., Gambatese, J.~A., 2010. Qualitative research: Application of
  the {Delphi} method to {CEM} research. Journal of Construction Engineering
  and Management 136~(1), 99--107.

\bibitem[{Hoffman and Gelman(2014)}]{hoffman2014}
Hoffman, M.~D., Gelman, A., 2014. The no-u-turn sampler: adaptively setting
  path lengths in hamiltonian monte carlo. Journal of Machine Learning Research
  15~(1), 1593--1623.

\bibitem[{Hollnagel(2012)}]{hollnagel2012fram}
Hollnagel, E., 2012. {FRAM}, the functional resonance analysis method:
  modelling complex socio-technical systems. Ashgate Publishing, Ltd.

\bibitem[{Kucukelbir et~al.(2015)Kucukelbir, Ranganath, Gelman, and
  Blei}]{kucukelbir2015}
Kucukelbir, A., Ranganath, R., Gelman, A., Blei, D., 2015. Automatic
  variational inference in stan. In: Cortes, C., Lawrence, N., Lee, D.,
  Sugiyama, M., Garnett, R. (Eds.), Advances in Neural Information Processing
  Systems. Vol.~28. Curran Associates, Inc., pp. 568--576.

\bibitem[{Lander et~al.(2011)Lander, Eisen, Stentz, Spanjer, Wendland, and
  Perry}]{lander2011near}
Lander, L., Eisen, E.~A., Stentz, T.~L., Spanjer, K.~J., Wendland, B.~E.,
  Perry, M.~J., 2011. Near-miss reporting system as an occupational injury
  preventive intervention in manufacturing. American journal of industrial
  medicine 54~(1), 40--48.

\bibitem[{Lawson(2018)}]{lawson2018}
Lawson, A.~B., 2018. Disease map reconstruction and relative risk assessment,
  third edition. Edition. CRC Press, Boca Raton, FL.

\bibitem[{Neal and Griffin(2004)}]{neal2004safety}
Neal, A., Griffin, M.~A., 2004. Safety climate and safety at work. In: Barling,
  J., Frone, M.~R. (Eds.), The psychology of workplace safety. American
  Psychological Association, pp. 15--34.

\bibitem[{Nocedal and Wright(2006)}]{nocedal2006}
Nocedal, J., Wright, S.~J., 2006. Quasi-Newton Methods. Springer New York, New
  York, NY.

\bibitem[{Paiva and Tewari(2022)}]{paiva2021}
Paiva, A.~R., Tewari, A., 2022. Methodology for testing and evaluation of
  safety analytics approaches. Safety Science 152, 105737.

\bibitem[{Pflug(2000)}]{pflug2000}
Pflug, G.~C., 2000. Some remarks on the value-at-risk and the conditional
  value-at-risk. In: Uryasev, S.~P. (Ed.), Probabilistic Constrained
  Optimization: Methodology and Applications. Springer US, Boston, MA, pp.
  272--281.

\bibitem[{Pinto et~al.(2011)Pinto, Nunes, and Ribeiro}]{pinto2011}
Pinto, A., Nunes, I.~L., Ribeiro, R.~A., 2011. Occupational risk assessment in
  construction industry -- overview and reflection. Safety Science 49~(5), 616
  -- 624.

\bibitem[{Poh et~al.(2018)Poh, Ubeynarayana, and Goh}]{poh2018375}
Poh, C.~Q., Ubeynarayana, C.~U., Goh, Y.~M., 2018. Safety leading indicators
  for construction sites: A machine learning approach. Automation in
  Construction 93, 375 -- 386.

\bibitem[{Ringen and Englund(2006)}]{ringen2006construction}
Ringen, K., Englund, A., 2006. The construction industry. Annals of the New
  York Academy of Sciences 1076~(1), 388--393.

\bibitem[{Sarkar et~al.(2020)Sarkar, Pramanik, Maiti, and
  Reniers}]{sarkar2020predicting}
Sarkar, S., Pramanik, A., Maiti, J., Reniers, G., 2020. Predicting and
  analyzing injury severity: A machine learning-based approach using
  class-imbalanced proactive and reactive data. Safety science 125, 104616.

\bibitem[{Sheehan et~al.(2016)Sheehan, Donohue, Shea, Cooper, and
  De~Cieri}]{sheehan2016leading}
Sheehan, C., Donohue, R., Shea, T., Cooper, B., De~Cieri, H., 2016. Leading and
  lagging indicators of occupational health and safety: The moderating role of
  safety leadership. Accident Analysis \& Prevention 92, 130--138.

\bibitem[{Smith and Jones(2014)}]{smith2014hurt}
Smith, R.~M., Jones, M., 2014. A hurt-based approach to safety. In: IPTC 2014:
  International Petroleum Technology Conference. European Association of
  Geoscientists \& Engineers.

\bibitem[{Tixier et~al.(2017)Tixier, Hallowell, and
  Rajagopalan}]{tixier2017riskModeling}
Tixier, A. J.-P., Hallowell, M.~R., Rajagopalan, B., 2017. Construction safety
  risk modeling and simulation. Risk Analysis 37~(10), 1917--1935.

\bibitem[{Tixier et~al.(2016{\natexlab{a}})Tixier, Hallowell, Rajagopalan, and
  Bowman}]{tixier2016application}
Tixier, A. J.-P., Hallowell, M.~R., Rajagopalan, B., Bowman, D.,
  2016{\natexlab{a}}. Application of machine learning to construction injury
  prediction. Automation in construction 69, 102--114.

\bibitem[{Tixier et~al.(2016{\natexlab{b}})Tixier, Hallowell, Rajagopalan, and
  Bowman}]{tixier2016automated}
Tixier, A. J.-P., Hallowell, M.~R., Rajagopalan, B., Bowman, D.,
  2016{\natexlab{b}}. Automated content analysis for construction safety: A
  natural language processing system to extract precursors and outcomes from
  unstructured injury reports. Automation in Construction 62, 45--56.

\bibitem[{Verma et~al.(2018)Verma, Chatterjee, Sarkar, and
  Maiti}]{verma2018data}
Verma, A., Chatterjee, S., Sarkar, S., Maiti, J., 2018. Data-driven mapping
  between proactive and reactive measures of occupational safety performance.
  In: Industrial Safety Management. Springer, pp. 53--63.

\bibitem[{Vose(2008)}]{vose2008}
Vose, D., 2008. Risk Analysis: A Quantitative Guide. Wiley.
\newline\urlprefix\url{https://books.google.com/books?id=9CaoAqaRcVwC}

\bibitem[{Wu et~al.(2011)Wu, Chang, Shu, Chen, and Wang}]{wu2011716}
Wu, T.-C., Chang, S.-H., Shu, C.-M., Chen, C.-T., Wang, C.-P., 2011. Safety
  leadership and safety performance in petrochemical industries: The mediating
  role of safety climate. Journal of Loss Prevention in the Process Industries
  24~(6), 716--721, papers Presented at the 2010 International Symposium of the
  Mary Kay O'Connor Process Safety Center.

\bibitem[{Wu et~al.(2008)Wu, Chen, and Li}]{wu2008correlation}
Wu, T.-C., Chen, C.-H., Li, C.-C., 2008. A correlation among safety leadership,
  safety climate and safety performance. Journal of Loss Prevention in the
  Process Industries 21~(3), 307--318.

\bibitem[{Yorio et~al.(2014)Yorio, Willmer, and Haight}]{yorio2014interpreting}
Yorio, P.~L., Willmer, D.~R., Haight, J.~M., 2014. Interpreting msha citations
  through the lens of occupational health and safety management systems:
  Investigating their impact on mine injuries and illnesses 2003--2010. Risk
  analysis 34~(8), 1538--1553.

\bibitem[{Zou et~al.(2017)Zou, Kiviniemi, and Jones}]{zou201766}
Zou, Y., Kiviniemi, A., Jones, S.~W., 2017. Retrieving similar cases for
  construction project risk management using natural language processing
  techniques. Automation in Construction 80, 66 -- 76.

\end{thebibliography}
